\documentclass[12pt,preprint]{aastex}

\newcommand{\kms}{km~s$^{-1}~$}
\newcommand{\kmsMpc}{km~s$^{-1}$~Mpc$^{-1}~$}

\shorttitle{SN~2002fk}
\shortauthors{Cartier et al.}

\begin{document}

\title{\bf Persistent C~II Absorption in the Normal Type Ia Supernova 2002fk\altaffilmark{*}}
\author{R\'egis Cartier\altaffilmark{1, 2}, Mario Hamuy\altaffilmark{2, 1}, Giuliano Pignata\altaffilmark{3, 1}, Francisco F\"orster\altaffilmark{4, 1},
Paula Zelaya\altaffilmark{1, 5}, Gaston Folatelli\altaffilmark{6}, Mark M.~Phillips\altaffilmark{7}, Nidia Morrell\altaffilmark{7},
Kevin Krisciunas\altaffilmark{8}, Nicholas B.~Suntzeff\altaffilmark{8}, Alejandro Clocchiatti\altaffilmark{5, 1}, Paolo Coppi\altaffilmark{9},
Carlos Contreras\altaffilmark{7}, Miguel Roth\altaffilmark{7}, Kathleen Koviak\altaffilmark{10}, Jos\'e Maza\altaffilmark{2},
Luis Gonz\'alez\altaffilmark{2}, Sergio Gonz\'alez\altaffilmark{7} and Leonor Huerta\altaffilmark{2}}
\altaffiltext{1}{Millennium Institute of Astrophysics, Casilla 36-D, Santiago, Chile}
\altaffiltext{2}{Departamento de Astronom\'ia, Universidad de Chile, Casilla 36-D, Santiago, Chile}
\altaffiltext{3}{Departamento Ciencias Fisicas, Universidad Andres Bello, Av. Rep\'ublica 252, Santiago, Chile}
\altaffiltext{4}{Center for Mathematical Modelling, Universidad de Chile, Avenida Blanco Encalada 2120, Piso 7, Santiago, Chile}
\altaffiltext{5}{Departamento de Astronom\'ia y Astrof\'isica, Pontificia Universidad Cat\'olica de Chile, Casilla 306, Santiago, Chile}
\altaffiltext{6}{Institute for the Physics and Mathematics of the Universe (IPMU), University of Tokyo, 5-1-5 Kashiwanoha, Kashiwa, Chiba 277-8583, Japan}
\altaffiltext{7}{Carnegie Institution of Washington, Las Campanas Observatory, Colina el Pino s/n, Casilla 601, Chile}
\altaffiltext{8}{Department of Physics and Astronomy, Texas A\&M University, 4242 TAMU, College Station, TX~77843, USA}
\altaffiltext{9}{Department of Astronomy, Yale University, New Haven, CT 06520-8101}
\altaffiltext{10}{Carnegie Institution of Washington, 813 Santa Barbara Street, Pasadena, CA~911901, USA}
\altaffiltext{*}{This paper includes data gathered with the $6.5$~m Magellan telescopes at Las Campanas Observatory, Chile}

\begin{abstract}

\noindent
We present well-sampled  $UBVRIJHK$ photometry of SN~2002fk starting 12 days before maximum light
through 122 days after peak brightness, along with a series of 15 optical spectra from -4 to +95 
days since maximum. Our observations show the presence of C~II lines in the early-time spectra of 
SN~2002fk, expanding at ~11,000~\kms and persisting until ~8 days past maximum light with a velocity 
of $\sim$9,000~\kms. SN~2002fk is characterized by a small velocity gradient of $\dot v_{Si~II}=26$~km~s$^{-1}$~day$^{-1}$,
possibly caused by an off-center explosion with the ignition region oriented towards the observer.
The connection between
viewing angle of an off-center explosion and the presence of C II in the early time spectrum suggests
that the observation of C~II could be also due to a viewing angle effect.
Adopting the Cepheid distance to NGC 1309 
we provide the first $H_{0}$ value based on near-IR measurements of a Type Ia supernova between
63.0$\pm${0.8}~($\pm$ 2.8 systematic) and 66.7$\pm${1.0}~($\pm$ 3.5 systematic) \kmsMpc, depending on 
the absolute magnitude/decline rate relationship adopted.
It appears that the near-IR yields somewhat lower (6-9\%) $H_0$ values than the optical.
It is essential to further examine this issue
by (1) expanding the sample of high-quality near-IR light curves of SNe in the Hubble flow, and
(2) increasing the number of nearby SNe with near-IR SN light curves and
precise Cepheid distances, which affords the promise to deliver a more precise determination of $H_0$.




\end{abstract}

\keywords{supernovae: general - supernovae: individual: SN~2002fk - cosmological parameters - cosmology: observations }

\section{INTRODUCTION}

Type Ia supernovae (SNe) have proven to be extremely useful as extragalactic distance indicators in the 
optical \citep{hamuy06,hicken09,fola10} thanks to the relation between peak luminosity and decline 
rate \citep{phillips93}, which allows the standardization of their luminosities, and in the near 
infrared \citep{wood08,kris04b,kris04c} where they behave almost as standard candles \citep{kattner12}.
Photometric observations of SNe Ia have delivered Hubble diagrams with unrivaled low dispersions (0.15 mag),
which permits one to determine precise values for fundamental cosmological parameters such as the Hubble
constant and the deceleration parameter. It was this technique that led in 1998 to the
surprising discovery that the Universe is in a state of accelerated expansion caused by a mysterious dark 
energy that comprises 70\% of the energy content of the Universe \citep{riess98,perl99}. Measuring the
equation of state parameter of the dark energy and its evolution with time constitutes one of the most
important astrophysical challenges today.

Since the discovery of the accelerated expansion, extensive surveys and follow-up programs have increased
the number of SNe Ia by more than an order of magnitude. Furthermore, the quality of the follow--up observations
(light curves, early--time optical spectra, near-infrared (near-IR) spectra, nebular spectroscopy, polarization) 
has significantly increased. As a consequence of the increase in the statistical samples available and the degree
of details observed, several indications of diversity have emerged within the family of SNe Ia that manifest in
differences in several observables, most notably, early time velocity gradients \citep{benetti04}, the presence
of unburned material, late-time nebular velocity shifts \citep{maeda10a, maeda10c}, line
polarization \citep{leonard05,wangwheeler08}, high-velocity components of spectral lines \citep{kasen03, thomas04, gerardy04, mazzali05},
and interstellar line strengths \citep{2011Sci...333..856S, 2012ApJ...752..101F, 2012ApJ...754L..21F, phillips13}.
At this point the use of SNe Ia for the determination of cosmological parameters and the equation of state 
of the dark energy is dominated by systematics \citep[see e.g.][]{kessler09, conley11}. Among the most important
sources of systematics are the extinction by dust in the SN host galaxies \citep[][and references therein]{fola10}
and our lack of understanding of the explosion mechanisms and the progenitor system
\citep[see e.g.][]{2000ARA&A..38..191H, 2012ApJ...750L..19R, maoz14}.

SNe Ia are believed to result from the nuclear burning of a C+O white dwarf (WD). Given that carbon is not 
produced in the explosion, it provides unique evidence of the original composition of the star before explosion. 
For this reason the amount and distribution of the unburned carbon material is fundamental to understand the 
burning process of the star. Until recently only a handful of normal SNe Ia have shown carbon. Thanks to 
large surveys that have discovered and followed--up SNe Ia from very early phases, carbon has been found in nearly 
30\% of the SNe before maximum light \citep{parrent11, thomas11b, fola12, silverman12b}, and even more frequently
within the class of very luminous and low expansion velocities objects thought to be the result of a 
super--Chandrasekhar--mass WD \citep{howell06, thomas07, yamanaka09, scalzo10, silverman10}.

Additional multi-wavelength and multi-epoch observations of SNe Ia will be required to establish a scenario 
connecting the observed diversity among SNe Ia. Here we present a contribution to this subject through extensive 
optical and near-IR observations for the normal Type Ia SN~2002fk obtained in the course of the 
``Carnegie Type II Supernova Survey'' (CATS, hereafter) \citep{hamuy09}. While our optical data are complementary
to the photometry published by the Lick Observatory Supernova Search \citep[LOSS, ][]{riess09b, ganesha10}, and 
by the Center for Astrophysics (CfA) Supernova Program \citep[CfA3 sample, ][]{hicken09}, our near-IR data provide
one of the earliest near-IR light curves for a Type Ia SN. The exceptional coverage of the maximum both in the
optical and the near-IR light curves together with the Cepheid distance recently measured by \citet{riess11, riess09a},
makes SN~2002fk one of the most suitable SNe Ia for calibration of the Hubble constant. The completeness of the 
light curves also affords the opportunity to test independent methods for the determination of the host galaxy
reddening using a mixture of \bv and V-near--IR colors. In addition, we present 15 optical spectra obtained 
from $-4$ to $+95$ days since maximum in the $B$-band, which are complemented by spectra obtained by the CfA 
Supernova Program \citep[]{blondin12} and the Berkeley Supernova Ia Program \citep[BSNIP, ][]{silverman12a}.
The spectra provide clear evidence for unburned carbon. SN~2002fk is a very well observed SNe Ia showing carbon 
lines in pre-maximum spectra and the only one where such lines persist several days past peak brightness. Thus, these 
observations provide important clues to our understanding of the explosion mechanisms for normal SNe Ia.

This paper is organized as follows: in section \ref{obs} we describe the photometric and spectroscopic observations 
and data reductions; in section \ref{lightcurves} we present the photometric data and the determination of reddening 
caused by dust; in section \ref{spectra} we show the spectroscopic evolution of SN~2002fk; in section \ref{h0_sec}
we calculate the Hubble constant using the photometry presented here complemented by LOSS and CfA3 photometry, and 
the distance to NGC~1309 \citep{riess11}; we discuss our results and present our 
conclusions in section \ref{discussion}. In an upcoming paper \citep{zelaya13} we will present spectropolarimetric 
observations of SN~2002fk.

\section{OBSERVATIONS AND DATA REDUCTION}
\label{obs}

\begin{figure}
\plotone{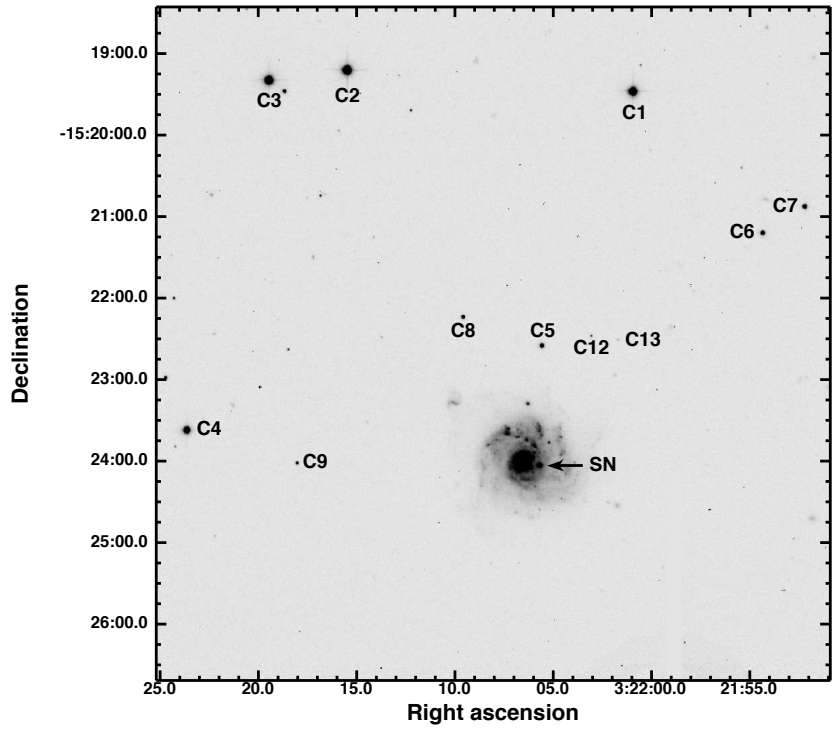}
\caption{Field of SN~2002fk observed with the Swope telescope. North is up and east is to the left. 
SN~2002fk is marked with an arrow and eleven stars of the local standard stars used to derive differential
photometry are labeled.}
\label{localseq_fig}
\end{figure}

SN~2002fk was discovered independently by \citet{kushida02} and by \citet{wang02} at $V \simeq 15.0$ mag on 2002 
September 17.7. The supernova was located at $\alpha$=03:22:05.71 and $\delta$=-15:24:03.2 (J$2000$), $3.6''$ 
south and $12.6''$ west of the host galaxy NGC 1309 (see Figure \ref{localseq_fig}). NGC 1309 is an SA(s)bc galaxy 
with a heliocentric recession velocity of 2,136 \kms obtained from neutral hydrogen line measurements \citep{kori04}.
\citet{ayani02} classified SN~2002fk as a Type Ia from an optical spectrum obtained on 2002 September 20.

\subsection{Instrumental Settings}

The optical imaging of SN~2002fk was obtained with UBVRI Johnson-Kron-Cousins filters using the Swope 1 m 
telescope, the Wide Field Re-imaging CCD Camera (WFCCD) on the 2.5 m du Pont telescope, and the Low-Dispersion 
Survey Spectrograph (LDSS-2) on the 6.5 m f/11 Magellan Clay telescope, all at Las Campanas Observatory (LCO).
A few additional optical images were obtained with the 0.9 m and 1.5 m telescopes at Cerro Tololo Inter-American 
Observatory (CTIO). For galaxy subtraction, we used images obtained with the Direct CCD Camera on the 2.5 m du Pont
telescope. Additional observations with the Panchromatic Robotic Optical Monitoring and Polarimetry Telescope (PROMPT) 
at CTIO were obtained for the single purpose of calibrating a photometric sequence of local standards around
SN~2002fk.

IR images of SN~2002fk were obtained with $JHK_s$ filters with the Swope telescope IR Camera (C40IRC) and the Wide 
Field IR Camera (WIRC) on the du Pont telescope. In Table \ref{Phot_tab} we summarize the instruments used for 
the optical and IR imaging.

Optical spectroscopy of SN~2002fk was obtained with five instruments: LDSS-2 and the B\&C Spectrograph on the LCO Baade 
telescope, the WFCCD and Modular Spectrograph on the LCO du Pont telescope, and FORS1-PMOS on the ESO (European Southern
Observatory) Very Large Telescope located on Cerro Paranal. In Table \ref{Spec_tab} we summarize the instruments used 
for the spectroscopic follow-up and their main characteristics. More details about these instruments are given 
by \citet{hamuy09}.

\subsection{Optical Photometry Reduction}
\label{obs_op}

We reduced the optical images using a custom IRAF\footnote[11]{IRAF is distributed by the National Optical Astronomy
Observatories, which are operated by the Association of Universities for Research in Astronomy, Inc., under cooperative 
agreement with the National Science Foundation.} script package, as described by \citet{hamuy09}. We processed the 
raw images through bias subtraction and flat fielding. As long as sky flats were available, we corrected small gradients
in dome flat illumination ($\thicksim 1 \%$). To reduce any possible galaxy contamination in our photometry we subtracted 
the host galaxy from our SN images using high quality template images obtained with the Direct CCD Camera on the du Pont
telescope after the SN had vanished on 2005 February 13. This instrument delivers a typical point-spread-function (PSF) 
of $0.7''$ (full width at half-maximum intensity), thus providing templates with better image quality than any of the 
SN+galaxy images. To secure high quality image subtractions, we employed the same $UBV(RI)_{KC}$ filter set of the Swope
telescope (which was used on the majority of our optical observations; see Table \ref{OpPhotometry_tab}). Differential PSF
photometry was made on the galaxy-subtracted images using the sequence of local standards (Figure \ref{localseq_fig}), 
calibrated against \citet{landolt92} photometric standards observed during several photometric nights. Systematic errors
introduced by using template images taken with the du Pont telescope are considered in Appendix \ref{cam_subtractions}.

The Landolt stars were employed to derive the following photometric transformations from instrumental to standard magnitudes,

\begin{equation}
U~=~u~+~ct_u~(u~-~b)~+~zp_u
\label{U_eq}
\end{equation}
\begin{equation}
B~=~b~+~ct_b~(b~-~v)~+~zp_b
\label{B_eq}
\end{equation}
\begin{equation}
V~=~v~+~ct_v~(v~-~i)~+~zp_v
\label{V_eq}
\end{equation}
\begin{equation}
R~=~r+~ct_r~(v~-~r)~+~zp_r
\label{R_eq}
\end{equation}
\begin{equation}
I~=~i+~ct_i~(v~-~i)~+~zp_i
\label{I_eq}
\end{equation}

In these equations $UBVRI$ (left-hand side) are the published magnitudes in the standard system \citet{landolt92},
$ubvri$ (right-hand side) correspond to the natural system magnitudes, $ct_j$ to the color term, and $zp_j$ to
the zero-point for filter $j (j=u,b,v,r,i)$\footnote[12]{In the case of the LDSS-2 instrument which does not have
an $I$ filter, we used $(b-v)$ instead of $(v-i)$ in equation \ref{V_eq}.}.

These transformations were then applied to 11 pre-selected stars in the field of the SN in order to establish a 
sequence of local standards, which is presented in Table \ref{OpSequence_tab}. In Appendix \ref{local_standards}
we perform a detailed analysis of our photometric calibration and a comparison with that obtained by LOSS \citep{ganesha10}
and CfA3 \citep{hicken09}.

Having established the photometric calibration of the SN~2002fk field, we carried out differential photometry between 
the SN and the local standards. For all five optical cameras we employed average color terms determined on multiple
nights (listed in Table \ref{coefficients_tab}), solving only for the photometric zero-points. The resulting $UBVRI$
magnitudes covering 130 days of evolution of SN~2002fk are summarized in Table \ref{OpPhotometry_tab}.

\subsection{Near-IR Photometry Reduction}

The vast majority (21/24) of the near-IR photometric observations were obtained with WIRC on the du Pont telescope.
WIRC has four different detectors. Usually only two of the detectors were used to observe. The observing strategy
consisted in taking five dithered images of the object with detector~1, while detector~2 was used to take $sky$ images.
Then the detectors were switched, i.e. the SN was observed on detector~2 while detector~1 was used to take $sky$ images.
The observations were reduced following standard procedures with a slighty modified version of the Carnegie Supernova
Project (CSP) WIRC pipeline \citep[written in IRAF, ][]{hamuy06, contreras10}. We had to modify the pipeline to
work without dark and bias because
these calibrations were not taken during the observations of SN~2002fk.
The pipeline first masks detectable sources in the sky frames and then combines them to create a final sky frame for
each detector. Then the pipeline subtracts the final sky image from the science images to get a bias/dark/sky subtracted
science frame. In order to correct for pixel--to--pixel gain variations we divide science images by a flat field
constructed from the subtraction of a flat-off from a flat-on image and normalize the result. For nights where flats
were not taken we used the best calibration images from the two closest nights. In addition to this, the pipeline 
masks cosmetic defects and saturated objects and combines the dithered frames from each detector to deliver a final
science image per detector.

Classic-Cam on the 6.5 m Baade telescope was used on three nights to obtain near-IR images. Since all calibrations
(dark, bias, flat) were taken with this instrument, we performed the standard data reduction. First we subtracted
the bias and the dark from all frames. Then we created a flat field for each night subtracting the flat--off from
the flat--on image and normalized the result. Finally, we corrected pixel--to--pixel variations dividing all science
images by the flat fields.

We obtained galaxy subtraction templates, using the WIRC/du~Pont configuration, on 2003 November 2 and 11 after the
SN had vanished. We perfomed differential PSF photometry of the SN in the WIRC galaxy--subtracted images with respect
to four near-IR local sequence stars around SN~2002fk. We calibrated these four local standards stars with respect 
to \citet{persson98} standards using aperture photometry. The Persson standard star images were taken close in time
and airmass to the SN field, so it was not necessary to correct for atmospheric extinction. Photometry of the local
sequence was obtained in four photometric nigths with WIRC. In Table \ref{IRSequence_tab} we present our $J H K_s$
photometry for the local sequence along with the 2MASS magnitudes for comparison. Within the uncertainties our 
photometry is in good agreement with the 2MASS values. This is an encouraging result considering that 
the \citet{persson98} system is not identical to the 2MASS photometric system \citep{skrutskie06}.

In the small field-of-view of Classic-Cam only star c5 was observed along with the SN. This precluded us from
subtracting the galaxy templates and performing PSF photometry, so we ended up measuring aperture photometry
of the SN with respect to c5 alone. Table \ref{IrPhotometry_tab} presents the resulting $JHK_S$ photometry
for SN~2002fk which covers 115 days of its evolution since discovery.

\subsection{Spectroscopy}

A total of 15 spectra of SN~2002fk were obtained, as summarized in Table \ref{Spec_tab}. We obtained five
spectra with the 6.5~m Magellan Baade telescope using LDSS-2. For these observations, a 300 line mm$^{-1}$
grism blazed at $5000$~\AA~ was employed, covering from 3600 to 9000~\AA~with a resolution of 14~\AA. 
Three spectra were observed with the Baade telescope using the B\&C Spectrograph. This instrument uses
a SITe $1752 \times 572$ pixel CCD with 300 line~mm$^{-1}$ grating blazed at $5000$~\AA. The wavelength range
obtained was 3200 to 9200~\AA~with a resolution of 7~\AA. Three times we used the du Pont telescope with the WFCCD
instrument in its spectroscopic long-slit mode. A 400~line~mm$^{-1}$ blue grism was employed, covering the wavelength
range from 3800 to 9200~\AA~with a resolution of 8~\AA . We also obtained one spectrum with the Las Campanas Modular
Spectrograph on the du Pont telescope, covering from 3790 to 7270~\AA~with a resolution of 7~\AA. With all four 
instruments (WFCCD, LDSS-2, B\&C, and Modular Spectrograph) we observed the SN and the standard stars with the slit
aligned along the parallactic angle \citep{filippenko82} to reduce the effects of atmospheric dispersion. Additional
information about the instruments and observing modes are presented in \citet{hamuy09}.

Three of our spectra of SN~2002fk were obtained in polarimetric mode with the FORS1 instrument on the ESO VLT UT3
telescope on 2002 October 1st, 5th, and 14th. For each of the three epochs observed, four exposures were taken with 
the retarder plate at position angles of 0, 22.5, 45 and 67.5 degrees. The exposure times were 720 sec on October 1st,
720 sec on October 5th, and 600 sec on October 14th. Grism GRIS\_300V was used with no order separation filter. This
provides a dispersion of 2.59~\AA~pix$^{-1}$, and the resulting wavelength range 3300--8500 \AA. The spectral resolution
with a 1$''$ slit was 12.25~\AA, which we measured using the [OI]~5577 sky emission line. Additional observations of a
flux standard, EG21, were performed at position angle 0 degrees to calibrate the flux level. We followed a standard data
reduction procedure using IRAF. A single intensity spectrum was obtained for each epoch, combining the four angle spectra
with two beams each, in the following way:

\begin{equation}{
F_{tot}=\frac{\sum\frac{spec_{i}}{specerr_{i}^2} } {\sum \frac{1}{specerr_{i}^2}}
}
\end{equation}

\noindent with i=0, 22, 45 and 67, where $spec_{i}$ is the sum of the two beams corresponding to ordinary and extraordinary rays,
at each angle, and $specerr_{i}$ is the quadratic sum of the errors per beam for each of the four angles. Here we present the
intensity spectra alone, and we postpone the spectropolarimetric reduction details and results for an upcoming paper \citep{zelaya13}.

Despite having observed the SN along the parallactic angle, we checked the spectrophotometric quality of our spectra
by convolving them with the standard $BV(RI)_{KC}$ bandpasses described by \citet{bessell90} and computing synthetic magnitudes.
In general, this comparison shows very good agreement within $0.01$-$0.03$ mag between the observed and synthetic colors, thus 
demonstrating the excellent quality of our spectra. Having checked the relative spectrophotometry we proceeded to apply
a low--order wavelength dependent correction to make all of our spectra fully consistent with the photometry.

\begin{figure}
\plotone{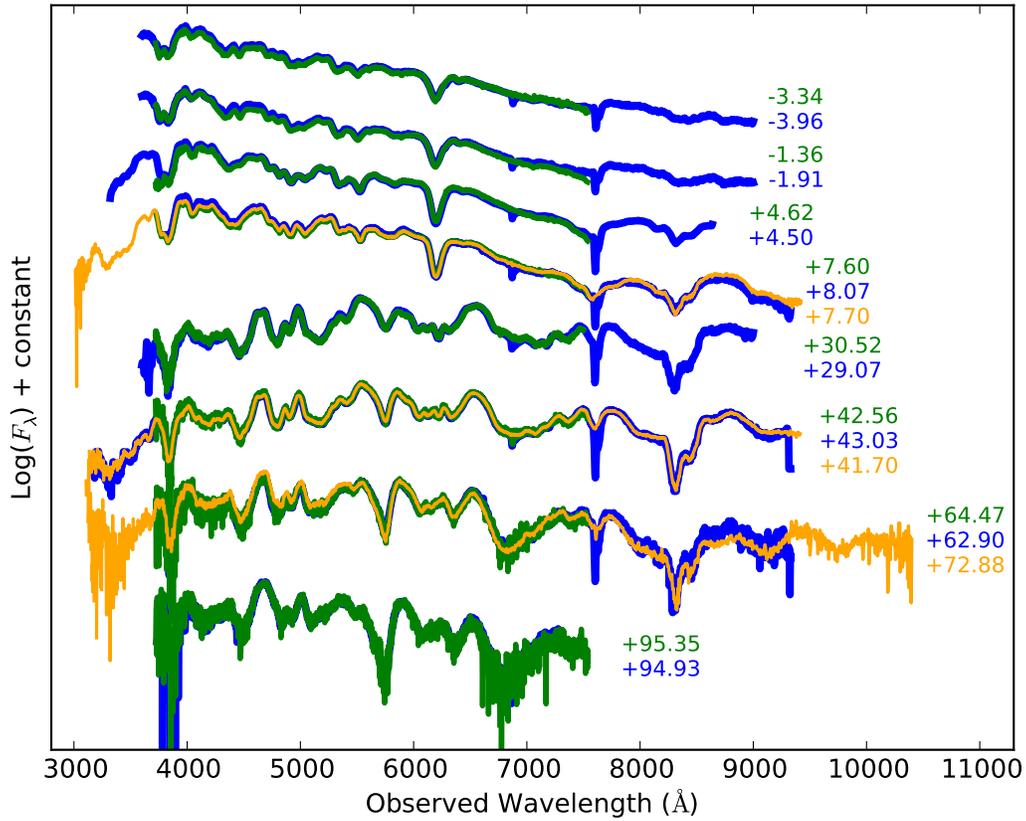}
\caption{Comparison between our spectra (blue), the CfA \citep[green,][]{blondin12}, and the BSNIP \citep[orange,][]{silverman12a}
spectra taken at similar epochs. The phase is indicated on the right for each spectrum.}
\label{comp_with_CfAnLOSS1_fig}
\end{figure}

Optical spectra of SN~2002fk were recently published by the CfA Program \citep[]{blondin12} and the Berkeley Supernova Ia 
Program \citep[BSNIP, ][]{silverman12a}. CfA obtained 23 spectra covering from $-3$ to $+120$ days since maximum in $B$ and 
BSNIP obtained five spectra from $+8$ to $+150$ days after $B$ maximum. In Figure \ref{comp_with_CfAnLOSS1_fig} we compare
our spectra (blue) taken with the CfA (green) and the BSNIP (orange) spectra observed within one day from ours. As can be seen,
the agreement is excellent, with only very small differences with the first two CfA spectra. All of these data will be used 
subsequently in our study.

\section{LIGHT CURVES OF SN~2002fk}
\label{lightcurves}

\subsection{Optical Light Curves}
\label{opl_sec3_1}

\begin{figure}
\plotone{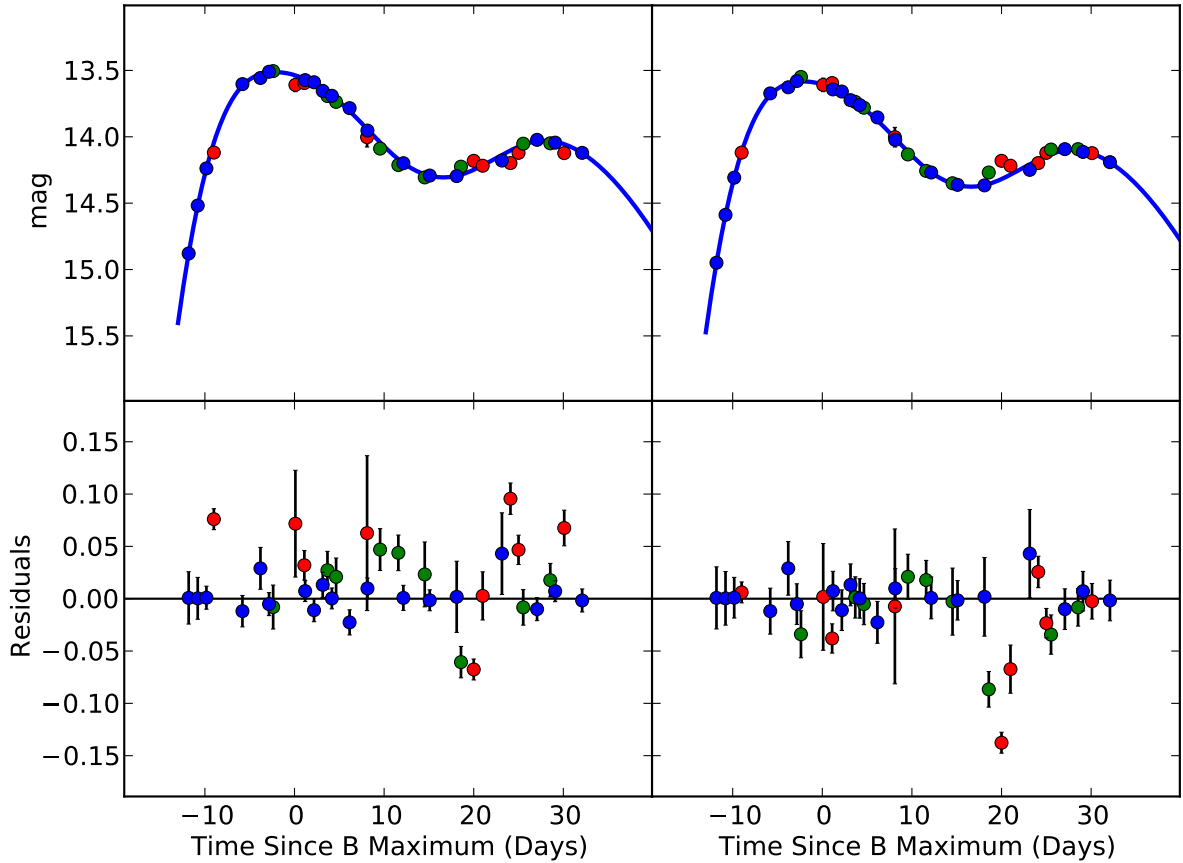}
\caption{Upper left. $I$-band light curve of SN~2002fk (red=CATS, blue=LOSS, green=CfA3) before applying the 
photometric offset found in the calibration of the local standards. Bottom left. Magnitude residuals from the
polynomial fit (blue line). Upper right. $I$-band light curve of SN~2002fk after applying the photometric offset caused
by systematic differences in the local standards, along with a polynomial fit (blue line). Bottom left. Magnitude
residuals from the polynomial fit.}
\label{LOSS_vs_CATS_CfA_I_fig}
\end{figure}

Optical photometry of SN 2002fk has been previously published by LOSS \citep[]{riess09b, ganesha10} and by the CfA
Supernova Program \citep[CfA3 sample, ][]{hicken09}. These data have good sampling, most notably the LOSS observations
that begin $11.8$ days before $B$-band maximum. These datasets are a valuable complement to our optical photometry. However,
since we have detected systematic differences between the CATS, LOSS, and CfA3 photometric calibrations it has proven 
necessary to apply offsets to the LOSS and CfA3 photometry in order to bring all measurements to the same CATS photometric
system, as described in Appendix \ref{local_standards}. In doing so, we added in quadrature to the LOSS and CfA3
uncertainties those from the photometric offsets. Adding these offsets (summarized in Table \ref{SysDiff_tab}) decrease
significantly the differences in the $R$- and $I$--band SN magnitudes among the three datasets (as illustrated in
Figure \ref{LOSS_vs_CATS_CfA_I_fig} for the $I$--band), but has no noticeable effects in the $B$- and $V$--bands.

\begin{figure}
\plotone{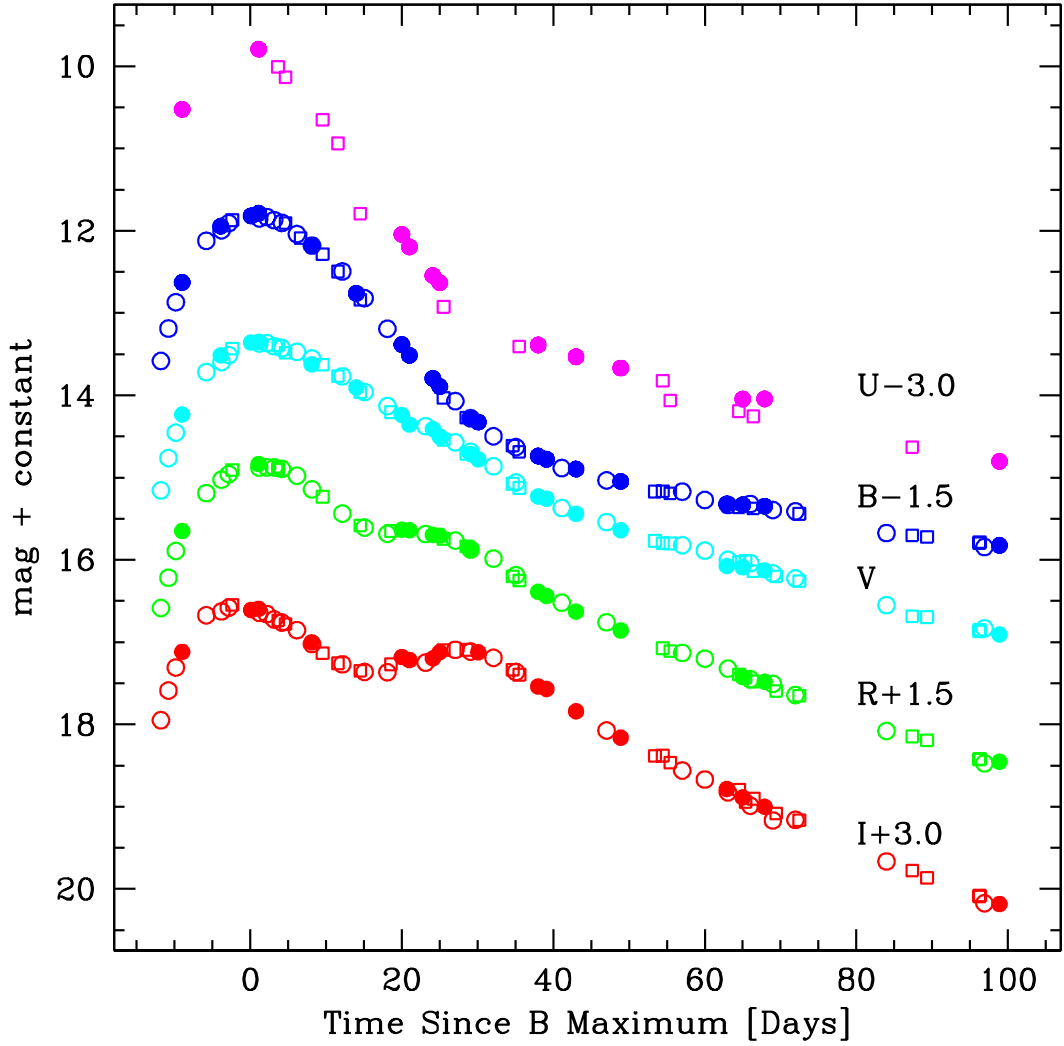}
\caption{Optical light curves of SN~2002fk. Filled circles are our observations, open circles correspond to LOSS measurements,
and open squares to CfA3 data. Photometry of LOSS and CfA3 were shifted by the offsets given in Table \ref{SysDiff_tab}
to bring them into our CATS photometric system (as discussed in Appendix \ref{local_standards}).}
\label{OpLC_fig}
\end{figure}

Figure \ref{OpLC_fig} presents the $UBVRI$ light curves of SN~2002fk, including CATS (filled circles) and the LOSS
(open circles) and CfA3 data (open squares) after applying the corresponding photometric offsets. A polynomial fit 
to the B-band light curve around maximum brightness yields $B_{max} = 13.30 \pm 0.03$ mag on JD $2,452,547.8 \pm 0.5$
and a decline rate parameter $\Delta m_{15}=1.02$ $\pm 0.04$ which is close to the fiducial value adopted 
by \citet{phillips99} and \citet{fola10} for SNe Ia. Polynomial fits to the other lightcurves yield 
$U_{max}=12.69 \pm 0.07$ mag on JD $2,452,546.0  \pm 1.5$, $V_{max}=13.37 \pm 0.03$ mag on JD $2,452,548.5  \pm 0.5$,
$R_{max}=13.36 \pm 0.03$ mag on JD $2,452,549.2  \pm 0.5$, and $I_{max}=13.57 \pm 0.03$ mag on JD $2,452,545.8  \pm 0.5$.
$V$ and $R$ peak roughly one day after $B$, whilst $U$ and $I$ peak two days before $B$. This is consistent with the typical
behavior of SNe Ia. Bright Type Ia SNe usually show a secondary maximum in $R$ and $I$, and in the near-IR bands. A polynomial
fit to the secondary peaks yields $R_{2^{nd}~max}=14.14 \pm 0.03$ on JD $2,452,568.3  \pm 0.5$ and $I_{2^{nd}~max}=14.08 \pm 0.03$
on JD $2,452,575.4  \pm 0.5$. Errors in peak magnitudes include the uncertainty due to differences in filter transmissions 
(S-corrections; see Appendix \ref{s_corrections}).

A small bump $\sim0.15$ mag is present in the $I$-band light curve of SN~2002fk located $\sim20$ days past maximum and 
$\sim7.5$ days before the secondary maximum. This is seen in Figure \ref{LOSS_vs_CATS_CfA_I_fig} which shows the individual 
$I$-band data points and the residuals from a polynomial fit. The left panels correspond to the original CATS, LOSS, and CfA3
data, while the right panels correspond to the same data after applying the photometric offsets. Although the bump in question
is supported by three photometric points (two from CATS and one from CfA3), each one with a 3-7 $\sigma$ significance, we cannot 
rule out the possibility that it is an artifact due to the combination of photometry taken with different filters and detectors.

\begin{figure}
\plotone{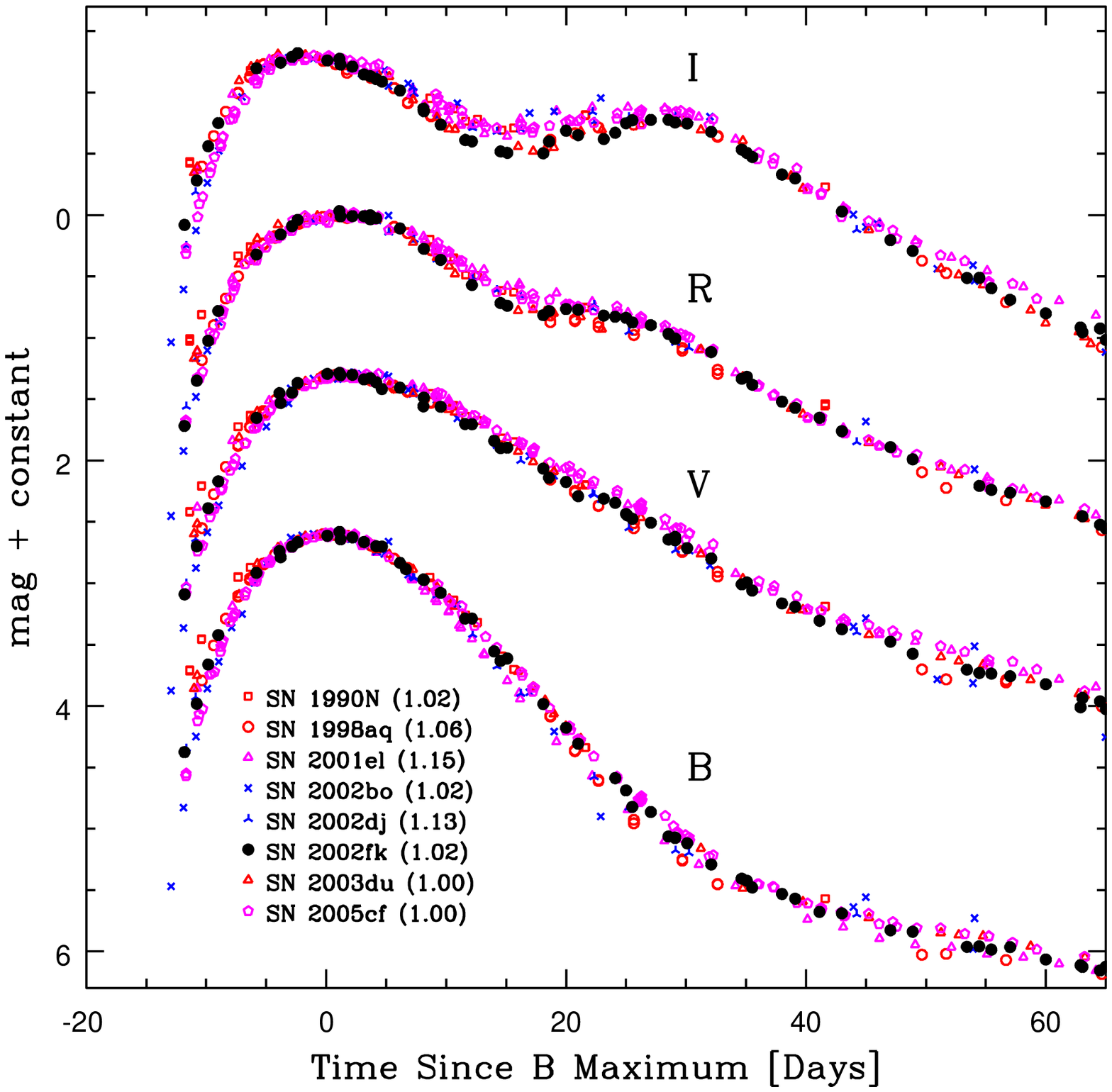}
\caption{$BVRI$ light curves of SN~1990N, SN~1998aq, SN~2001el, SN~2002bo, SN~2002dj, SN~2002fk, SN~2003du, and SN~2005cf,
\citep{lira98, riess05, kris03, benetti04,pignata08,stanishev07,wang09} selected for their similar decline rates, normalized
to peak brightness. $\Delta m 15$ is given in parenthesis for each SN.}
\label{Op_Comp_fig}
\end{figure}

In Figure \ref{Op_Comp_fig} we compare the $BVRI$ light-curves of SN~2002fk and other SNe with similar $\Delta m 15$. Despite 
their similar decline rates, we notice small yet real departures from a commom photometric behavior, namely: 1) a dispersion in 
the rise branch, where SN~2002bo, SN~2002dj, and SN~2005cf show faster rise times, in agreement to what was found by \citet{pignata08}
and \citet{ganesha11} where high-velocity gradient objects \citep[as defined by][] {benetti05} generally show shorter rise times; 
2) the small bump $ \sim 0.15~mag$ in the $I$--band light curve of SN~2002fk that peaks $20$ days past maximum in $B$;
3) small differences in the time and amplitude of the secondary peaks in $R$ and $I$ \citep[see also][]{kris01, fola10, burns11}.

\subsection{Near-IR Light Curves}

\begin{figure}
\plotone{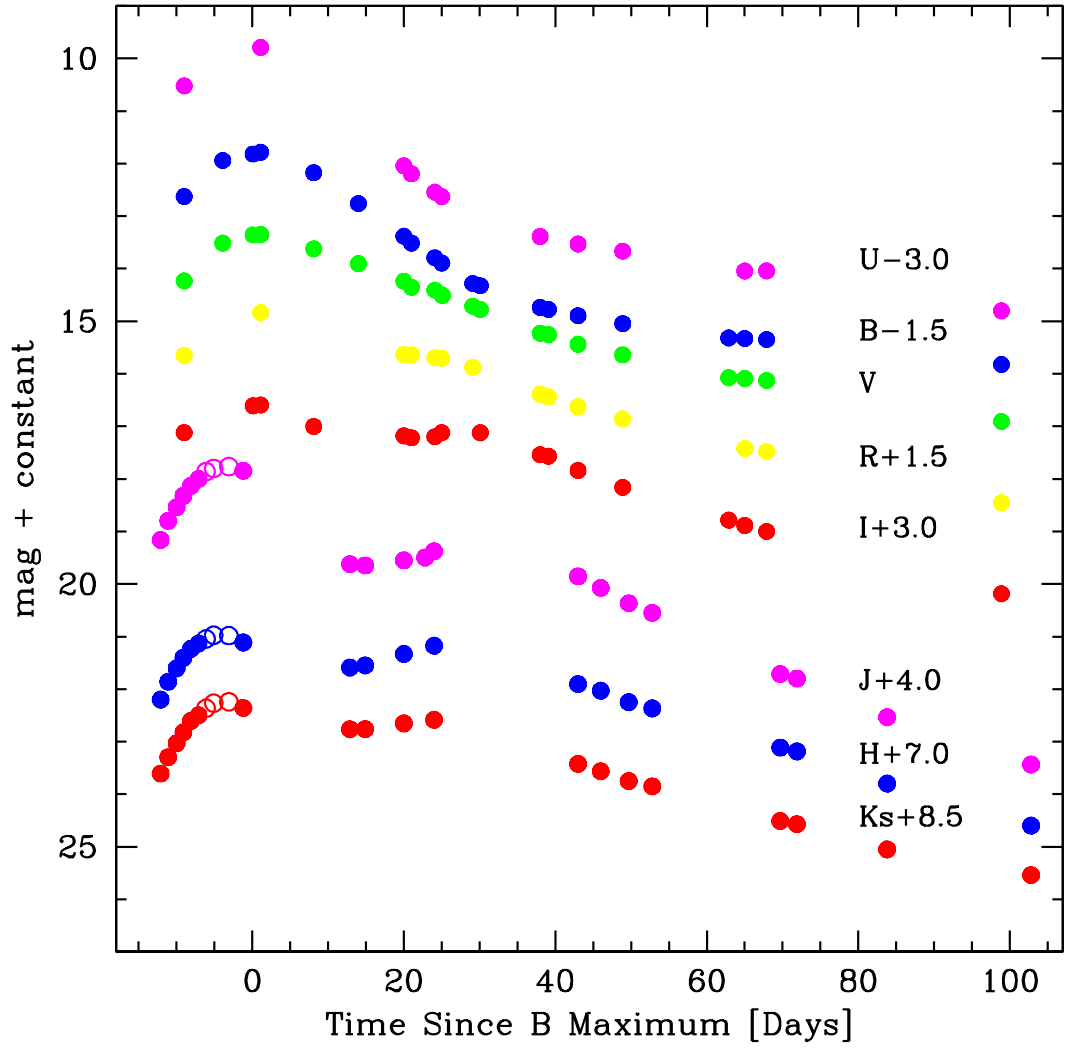}
\caption{Our optical and Near-IR photometry of SN~2002fk. In $JHK$ bands open symbols are data obtained with Classic-Cam, whilst 
filled symbols correspond to WIRC observations.}
\label{IRLC_fig}
\end{figure}

Figure \ref{IRLC_fig} shows our optical and near-IR light curves of SN~2002fk. Our near-IR observations begin 12 days before $B$-maximum,
among the earliest $JHK$ photometry ever obtained for a SN Ia, and extend through 102 days past maximum which is exceptional for
a SN Ia \citep{kris04b,wood08,fola10}. The photometric coverage is remarkably good before and close to peak brightness, through 
the near-IR minimum, and at late times. The second near-IR peak, on the other hand, is not so well sampled. Polynomial fits to
the IR light curves around maximum brightness yield $J_{max}=13.76 \pm 0.02$ on JD $2,452,544.3  \pm 0.5$, $H_{max}=13.98 \pm 0.02$
on JD $2,452,543.5  \pm 0.5$, and $K_{max}=13.76 \pm 0.02$ on JD $2,452,544.2  \pm 0.5$.

\begin{figure}
\plotone{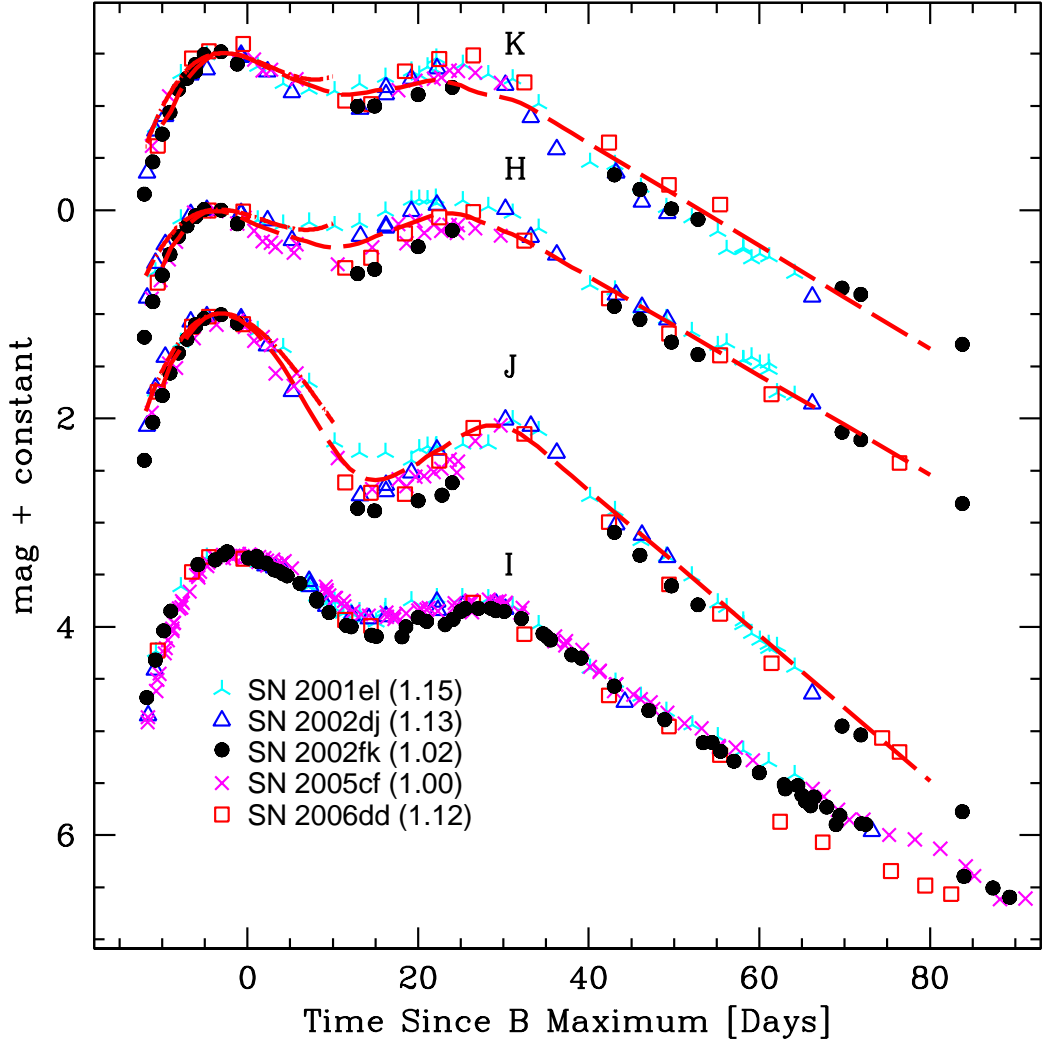}
\caption{$I J H K$ light curves of SN~2001el (cyan), SN~2002dj (blue), SN~2002fk (black), SN~2005cf (magenta), and SN~2006dd (red),
selected for their similar decline rates, normalized to peak brightness. We chose to compare the templates of \citet{kris04b} 
(dotted-dashed line) and \citet{wood08} (dashed line). $\Delta m 15$ is given in parenthesis for each SN.}
\label{IR_Comp_fig}
\end{figure}

In Figure \ref{IR_Comp_fig} we compare the $IJHK$-band light curves of SN~2002fk, SN~2001el \citep{kris03}, SN~2002dj \citep{pignata08},
SN~2005cf \citep{wang09}, and SN~2006dd \citep{stritzinger10} normalized to peak magnitudes in each band. These are well-studied SNe
with very complete optical and near-IR light curves and with similar $\Delta m_{15}$ values. We also show in dotted-dashed lines
the templates from \citet{kris04b} and in dashed lines the templates from \citet{wood08}. In all bands SN~2002fk has a faster rise
to peak than both templates. After maximum it is evident that SN~2002fk has a more pronounced minimum in all $IJHK$ bands compared
with the other SNe and a deeper minimum than the \citet{wood08} and \citet{kris04b} templates. Although the secondary peak was not
covered by our observations in $JHK$, the observations suggest a well defined secondary maximum, clearly distinguishable from the first
peak and from the minimum between the peaks. The differences in the light curves become much smaller around the secondary peak, after
which all five SNe show good agreement in their decline rates.

\citet{kasen06} showed that the near--IR secondary maximum is related to the ionization evolution of the iron group elements,
in particular to the transition of the iron/cobalt gas from double to singly ionized. In his scenario the mixing of $^{56}$Ni
outward increases the velocity extent of the iron core. This hastens the ocurrence of the near--IR secondary maximum. In 
SN~2002fk the secondary peak does not occur earlier than in the remaining SNe, so a large mixing of $^{56}$Ni is unlikely.
Higher metallicity progenitors should produce a higher fraction of stable iron group material \citep{timmes03}. \citet{kasen06}
showed that a large amount of iron group material would cause the recombination wave to reach the iron--rich layers relatively
earlier, thus producing an earlier secondary maximum. Given that in SN~2002fk the secondary peak was not observed to occur earlier,
greater than normal progenitor metallicity is also unlikely. In his models, \citet{kasen06} presented evidence that a higher contrast
between the peaks and the minima is expected in hotter SNe (i.e. those with relatively greater $^{56}$Ni mass) because the
transition wave from double to single ionized iron takes a longer time to reach the iron core. Perhaps then, the deep minimum
observed in SN~2002fk is produced by a higher than normal ejecta temperature, and the $I$-band bump $\sim$20 days past maximum
could be due to asymmetries in the distribution of iron group elements.




\subsection{Reddening Determination}
\label{reddening}

Given the early and complete optical and near-IR light curves of SN~2002fk we can estimate the amount of dust reddening
(Galactic + host galaxy) using different colors, assuming the standard Galactic extinction law of \citet{cardelli89}.

Firstly, we estimate the reddening towards SN~2002fk from its late-time $(B-V)$ colors using the standard Lira--Phillips
relation \citep{lira96, phillips99}, which yields $E(B-V)_{Lira} = 0.066 \pm 0.019$. The slope of SN~2002fk is in good
agreement with the Lira relation within the dispersion. Secondly, we use the near maximum optical colors versus decline
rate relations \citep{phillips99}, which yields $E(B-V)_{B_{max}-V_{max}} = 0.009 \pm 0.044$ and $E(B-V)_{V_{max} - I_{max}} = 0.082 \pm 0.038$.
Thirdly, we estimate the reddening using the $V-$ near-IR colors by minimizing $\chi^2$ between the observed $V-H$ and $V-K$
colors with respect to the \citet{kris04b} zero-reddening templates. From these fits we obtain $E(B-V)_{V-H}= 0.014 \pm 0.037$
and $E(B-V)_{V-K} = 0.075 \pm 0.036$. Within the uncertainties these results are consistent with the optical methods.

In table \ref{Reddening_tab} we summarize our results. The weighted mean of the reddening using these five different methods
is $E(B-V)=0.057$ with a dispersion of $0.036$ mag. This value is close to the measured Galactic reddening towards SN 2002fk,
namely $E(B-V)=0.035$ mag \citep[][]{schlafly11}, which implies that the host galaxy reddening was low. As shown 
by \citet{phillips13}, the equivalent width of Na I D is not a reliable indicator of host dust extinction in SNe Ia.
According to these authors, only the absence of detectable Na I D absorption in a high signal-to-noise ratio spectrum
can be used to infer low dust extinction. 
The fact that we detect a weak (0.018 \AA) Na I D absorption in our highest
signal-to-noise spectra at the redshift of the host galaxy is consistent with the small dust reddening inferred from the SN colors.

\subsection{Optical Colors}
\label{opcolors_sec}

\begin{figure}
\plotone{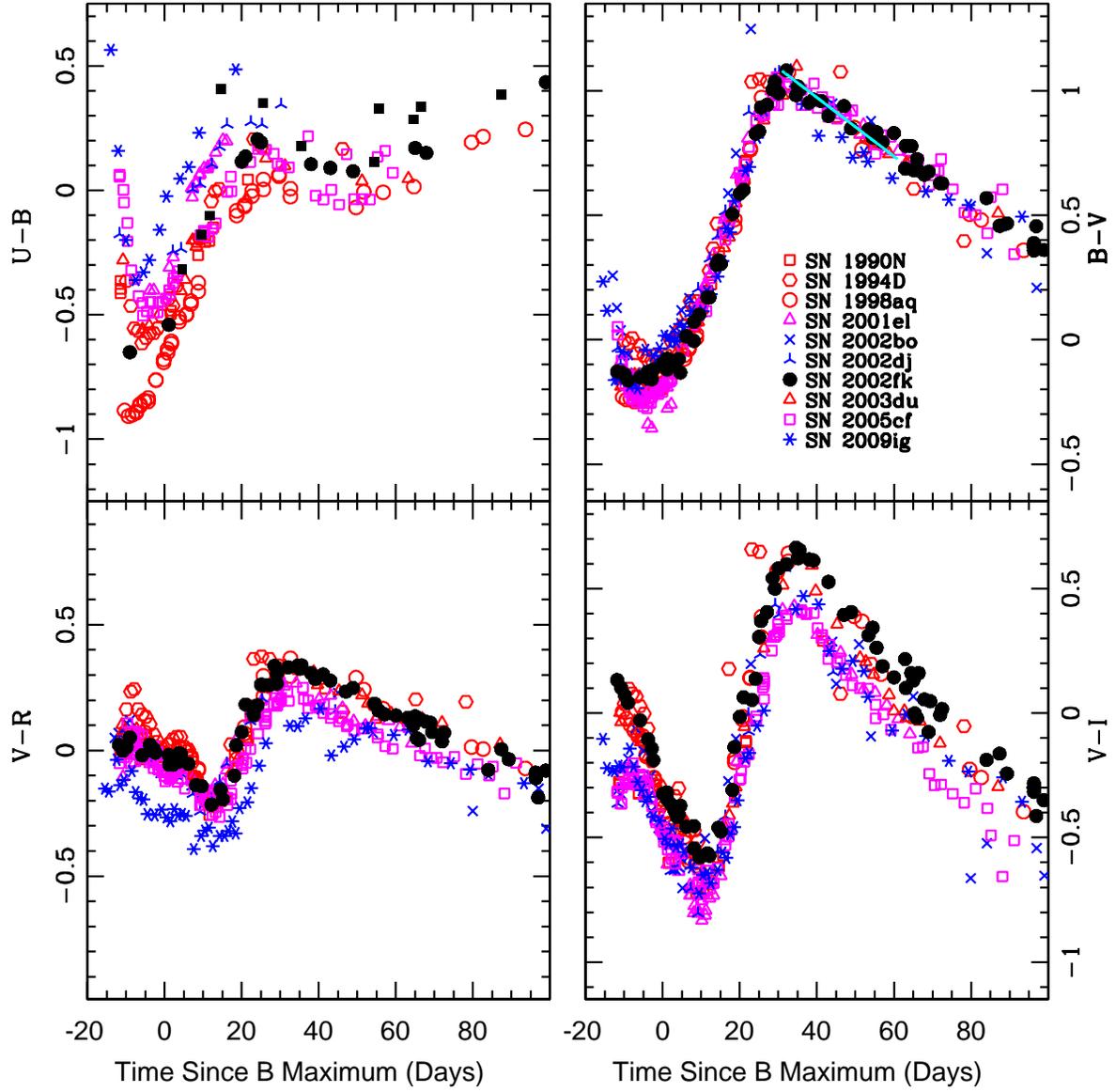}
\caption{Optical colors of SN~2002fk (in black). For comparison, in red are shown the LVG SN~1990N, SN~1994D, SN~1998aq, and SN~2003du.
In magenta are shown the LVG SN~2001el and SN~2005cf with strong HV Ca~II near maximum, and in blue the HVG SN~2002bo, SN~2002dj and SN~2009ig.}
\label{OpColors_fig}
\end{figure}

Figure \ref{OpColors_fig} shows dust corrected color curves of well observed ``normal'' SNe Ia with similar $\Delta m_{15}$
to SN~2002fk. We dereddened the color curves using the recalibrated Galactic extinction maps of \citet{schlegel98} presented by
\citet{schlafly11} and the host galaxy reddening derived from the Lira--Phillips relation \citep{lira96, phillips99}. In all cases
we transformed reddening into extinction at different filters using the standard Galactic extinction law of \citet{cardelli89} with
$R_{V}=3.1$, with the warning that the use of $R_{V}=3.1$ can introduce systematic offsets for heavily reddened SNe like 
SN~2002bo \citep{phillips13}, with the effect being largest in $V-I$. In red are shown SN~1990N \citep{lira98}, SN~1994D \citep{patat96},
SN~1998aq \citep{riess05}, SN~2003du \citep{stanishev07} which are LVG SNe.\footnote[13]{\citet{benetti05} separated SNe Ia into three 
groups: high-velocity gradient (HVG) objects, consisting of SNe with $\dot v_{Si~II} \gtrsim 70$~km~s$^{-1}$~day$^{-1}$ and 
$\Delta m 15 \lesssim 1.5$ mag, low-velocity gradient (LVG) objects, consisting of SNe with $\dot v_{Si~II} \lesssim 70$~km~s$^{-1}$~day$^{-1}$
and $\Delta m 15 \lesssim 1.5$, and FAINT objects with $\Delta m 15 \gtrsim 1.5$.} In magenta we show SN~2001el \citep{kris03} and
SN~2005cf \citep{wang09} both of which are LVG with strong high velocity Ca~II lines near maximum light. Finally in blue are SN~2002dj \citep{pignata08},
SN~2002bo \citep{benetti04}, and SN~2009ig \citep{foley12a} which are HVG SNe.

In the upper left panel of Figure \ref{OpColors_fig} we show the $U-B$ color curves. Overall, we see an evident dispersion amounting
to several tenths of a magnitude, despite all SNe having been corrected for dust reddening. We see a progression in color before and
near maximum, where the LVG SNe display the bluest colors, the LVG SNe with strong high-velocity (HV) Ca~II lines have intermediate colors,
and the HVG SNe are the reddest. What is more remarkable is the difference in shape between the two most complete $U-B$ color curves.
While SN~1998aq grows steadily redder in $U-B$ during the first 30 days, SN~2005cf shows a dramatic blueing at early epochs followed
by an upturn seven days before maximum. The evolution of SN~2002fk is similar to SN~1994D, evolving with a linear trend from very blue
$U-B$ colors at pre-maximum epochs to redder colors at about 30 days past maximum.  It is important to note at this point, that even
though the dispersion in $U-B$ colors can be significantly affected by systematic errors in dust reddening corrections, the differences
in shape cannot be accounted by this effect. Although filter transmission variations can reach a few tenths of a magnitude in the $U$-band,
the differences observed at early epochs reach even one magnitude, so we believe they are real and not a result of filter mismatches
or sky transmission.

The upper right panel of Figure \ref{OpColors_fig} shows the $B-V$ colors. Overall, the color evolution of SN~2002fk is that of a normal
Type Ia. The four SNe with the earliest observations (SN~1994D, SN~2002bo, SN~2005cf, SN~2003du, and SN~2009ig) all show clear evidence
for un upturn around 5 days before maximum that indicates an evolution from red to blue at the very early epochs. Even though the
sample is small, this seems to be a generic feature of SNe Ia. This behavior was already noted by \citet{hamuy91} in SN~1981D,
based on early-time observations obtained with photographic plates (see their Fig. 5). Promptly after maximum all SNe grow redder
for a period of $\sim$30 days, after which they all grow bluer which corresponds to the linear tail of the optical light curves.
As a reference we show with solid line the Lira law (in cyan). Within the uncertainties it appears that all SNe share a common
slope at these epochs, but we observe small systematic offsets of $\sim$0.08 mag for some individual objects. Considering that
the use of different instrumental bandpasses can introduce systematic errors $\sim$0.05 mag at these epochs \citep{stritzinger02},
it appears that there are some small yet significant intrinsic color differences among SNe Ia during the phase corresponding to the
Lira law.

The $V-R$ colors are shown in the lower left panel of Figure \ref{OpColors_fig}. The evolution of SN~2002fk has a normal shape
compared to the other SNe. However, we see an intrinsic scatter at the level of 0.4 mag among the different objects, with
SN~2002fk and SN~1998aq being among the reddest objects. SN~2005cf and SN~2001el, which are LVG objects with strong HV Ca~II
lines, are in the middle of the distribution, while the HVG objects are among the bluest objects. Note that this is the
opposite trend that we observe in $U-B$. This apparent gradient in $V-R$ color is specially noticeable 10 days before 
B maximum. At later times this gradient still seems to be present, but there is more scatter in the photometry.

In the lower right of Figure \ref{OpColors_fig} we show the $V-I$ color curves. We see evidence for a divergent behaviour at
the very earliest epochs. We can separate the SNe in two groups before maximum; the SN~2002fk-like SNe (LVG SNe without
strong Ca~II before maximum) showing a steady evolution from red to blue; the SN~2005cf-like group (LVG SNe with strong Ca~II lines);
and the HVG SNe, that show a downturn at day $\sim$-7. At about ten days before maximum, the former are nearly 0.3 mag redder
then the latter. An interesting case is SN~2009ig which belongs to the HVG group \citep{foley12a} and is slightly redder 
than SN~2005cf, but at $\sim$-15 days shows redder colors than the remaining SN~2005cf-like objects. The early-time
dichotomy observed in $V-I$ is reminiscent of what happens in $U-B$. Interestingly, both the $U$ and $I$ bands share
the presence of Ca~II lines, namely the Ca II H\&K lines at 3951~\AA~and the Ca IR triplet at 8579~\AA. However,
as demostrated by \citet{cartier11} the strength of the Ca~II lines cannot completely explain the dichotomy observed in
$U-B$ and $V-I$. About 10 days past $B$ maximum all SNe grow redder through 40 days after peak. Later on, all SNe grow bluer.
Although the slope is similar we observe significant offsets that exceed the potential systematic errors due to the use of different
instrumental bandpasses \citep{stritzinger02}.

\subsection{$V~-$~near-IR Colors}

\begin{figure}
\plotone{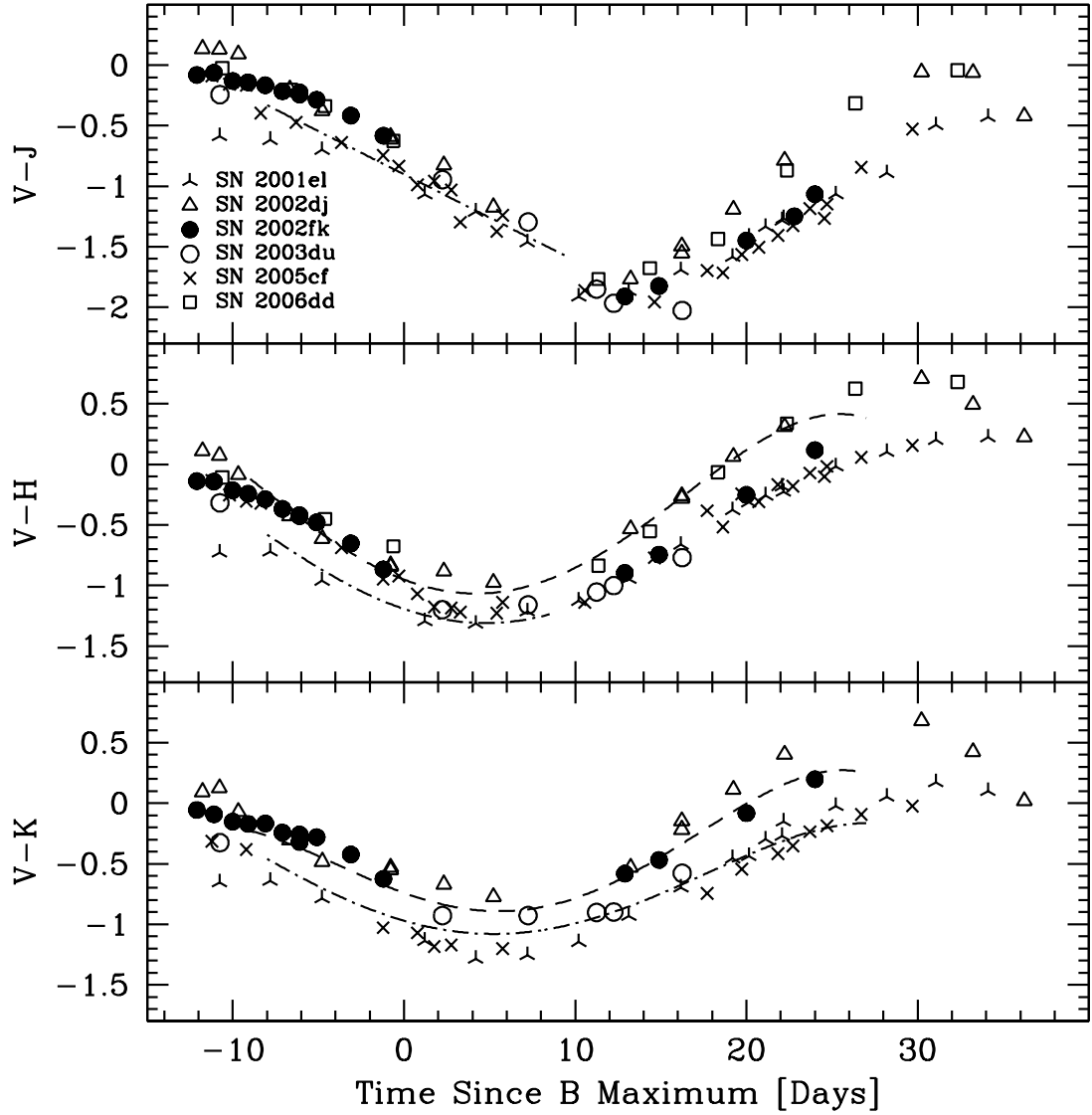}
\caption{$V~-$~near-IR colors of SN~2002fk and the well sampled SNe Ia of similar $\Delta m15$: SN~2001el, SN~2002dj, SN~2003du, SN~2005cf
and SN~2006dd. Dashed line are $V~-$~near-IR loci for mid-range decliners and dotted-dashed line are $V~-$~near-IR loci for slow decliners 
\citep{kris04b}.}
\label{IRColors_fig}
\end{figure}

Figure \ref{IRColors_fig} shows the dereddened $V~-$~near-IR colors of SN~2002fk. For comparison we show the $V~-$~near-IR
color curves of SN~2001el \citep{kris03}, SN~2002dj \citep{pignata08}, SN~2003du \citep{stanishev07}, SN~2005cf \citep{wang09},
and SN~2006dd \citep{stritzinger10}. These are well studied SNe with similar $\Delta m_{15}$, early, and complete near-IR
light curves.  As in Section \ref{opcolors_sec}, we corrected the colors making use of the Galactic extinction maps 
of \citet{schlafly11}, the Lira--Phillips relation to estimate host galaxy reddening, and the \citet{cardelli89}
extinction law to transform from color excess to extinction in specific bandpasses. We caution the reader that our
choice of $R_{V}=3.1$ may introduce significant offsets in the $V$--near-IR colors of heavily reddened objects with a lower
value of $R_{V}$.

We show in Figure \ref{IRColors_fig} the $V~-$~near-IR templates of \citet{kris04b}. In dotted-dashed lines are shown the loci
for slow decliners while dashed lines correspond to the loci of mid-range decliners. In the top panel we show the $V-J$ colors.
SN~2002fk declines from blue to red at a similar rate as the locus of slow decliners, but is offset by $\sim$0.2 mag to
redder colors. At early epochs SN~2002fk is similar to SN~2002dj, SN~2003du, and SN~2006dd. After ten days since maximum
SN~2002fk is similar to SN~2001el, and SN~2005cf.

The middle panel of Figure \ref{IRColors_fig} shows the $V-H$ color curves. Here SN~2002fk fits the locus of the mid-range
decliners at early epochs but after ten days SN~2002fk is $\sim$0.2 mag bluer than the template. At early times SN~2002fk
has similar colors to SN~2002dj, SN~2003du, SN~2005cf, and SN~2006dd. SN~2002dj and SN~2006dd resemble the mid-range
decliners template. After 10 days since maximum, the evolution of SN~2002fk continues to be similar to SN~2005cf.

The $V-K$ color curve is shown in the botton panel of Figure \ref{IRColors_fig}. SN~2002fk matches the template quite well,
although at early epochs it is slighty redder than the locus of the mid-range decliners. SN~2002dj is very similar to SN~2002fk,
whereas SN~2001el, SN~2003du, and SN~2005cf show colors that better fit the locus of the slow decliners.

\section{OPTICAL SPECTRA}
\label{spectra}

\begin{figure}
\plotone{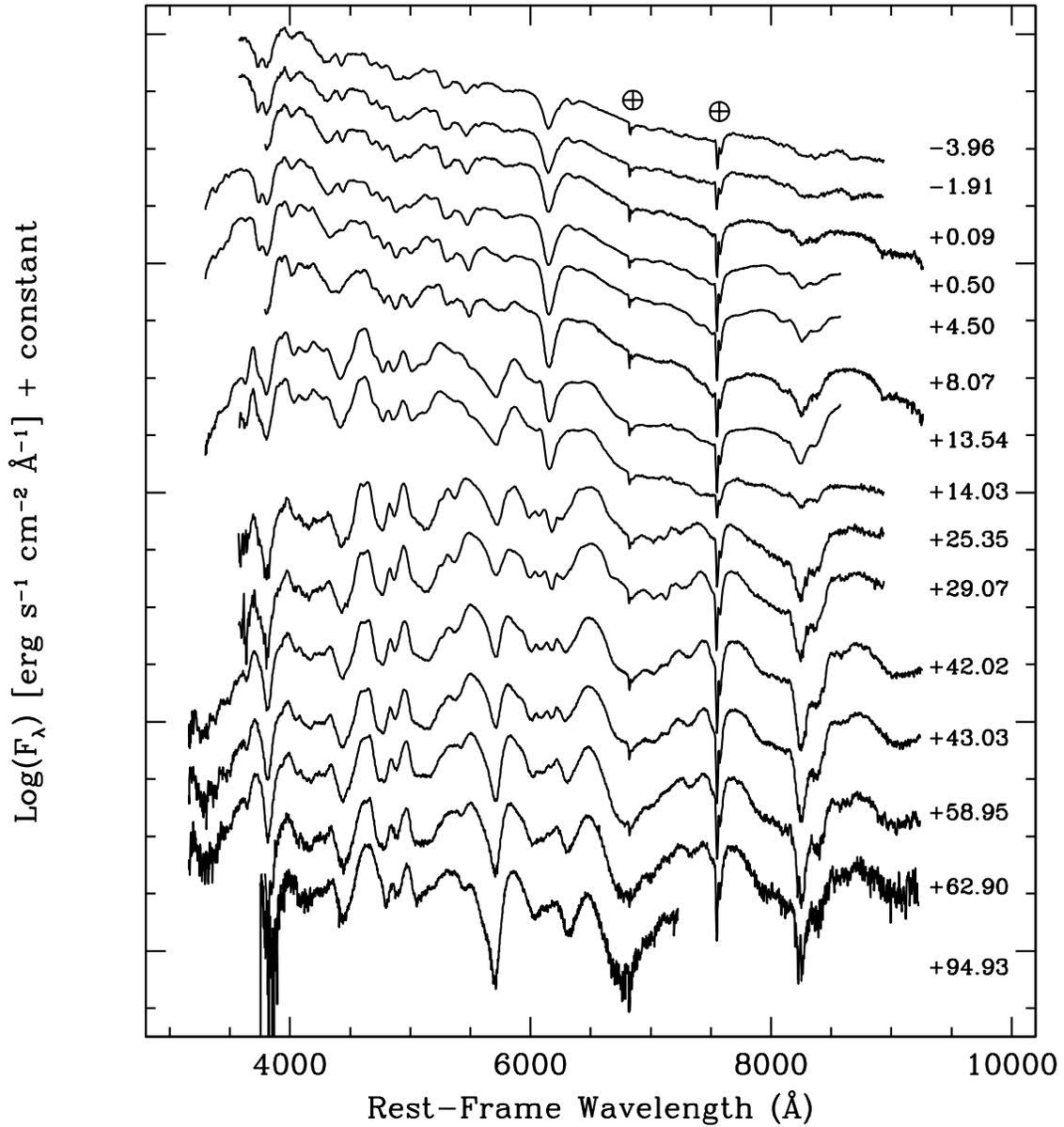}
\caption{Optical spectral evolution of SN~2002fk. The spectra have been corrected for the redshift of the host galaxy
(2,137 km s$^{-1}$) and have been shifted vertically for clarity. The labels on the right mark the days since
B-band maximum.}
\label{spectra_fig}
\end{figure}

Figure \ref{spectra_fig} shows 15 optical spectra of SN~2002fk that span from four days before B maximum to 95 days past peak.
Near maximum, SN~2002fk shows the characteristic Si~II $\lambda 6355$ absorption, Ca~II H\&K lines, the Ca~II near-IR triplet,
the W-shaped S~II lines, and Si~III $\lambda 4560$. \citet{nugent95} found a correlation between peak luminosity and {\it R}(Si~II),
the ratio of the intensities of Si~II $\lambda 5972$ to Si~II $\lambda 6355$ lines near maximum, which is interpreted as a
temperature effect. For SN~2002fk we measure {\it R}(Si~II) = 0.189 $\pm$ 0.002, corresponding to a bright SN with a hot 
ejecta \citep[see section \ref{h0_sec},][]{nugent95}. Further evidence that SN~2002fk was hotter than normal can be drawn 
from the blue \bv colors, and from the Si~III $\lambda 4560$ and $\lambda 5740$ absorption lines before maximum, which
are usually present in the early spectra of bright SNe.

A relatively weak yet evident absorption is present at $\sim 6400$ \AA, which we identify as C~II $\lambda 6580$. This line
was previously identified by \citet{blondin12} in a spectrum taken $-3.34$ days since maximum. Support for this identification
comes from two additional lines at $\sim 4600$ \AA~and $\sim 7000$ \AA~which can also be attributed to C~II. Remarkably, the C~II
$\lambda 6580$ feature persists until several days past maximum, which is unprecedented in normal SNe~Ia
\citep[see][]{parrent11,thomas11b,fola12,silverman12b}. Otherwise, the post-maximum evolution of SN~2002fk is consistent with
that of a Branch-normal Type Ia SNe (see section \ref{tempevol_sec} below).

\subsection{Temporal Evolution of the Spectra}
\label{tempevol_sec}

\begin{figure}
\plotone{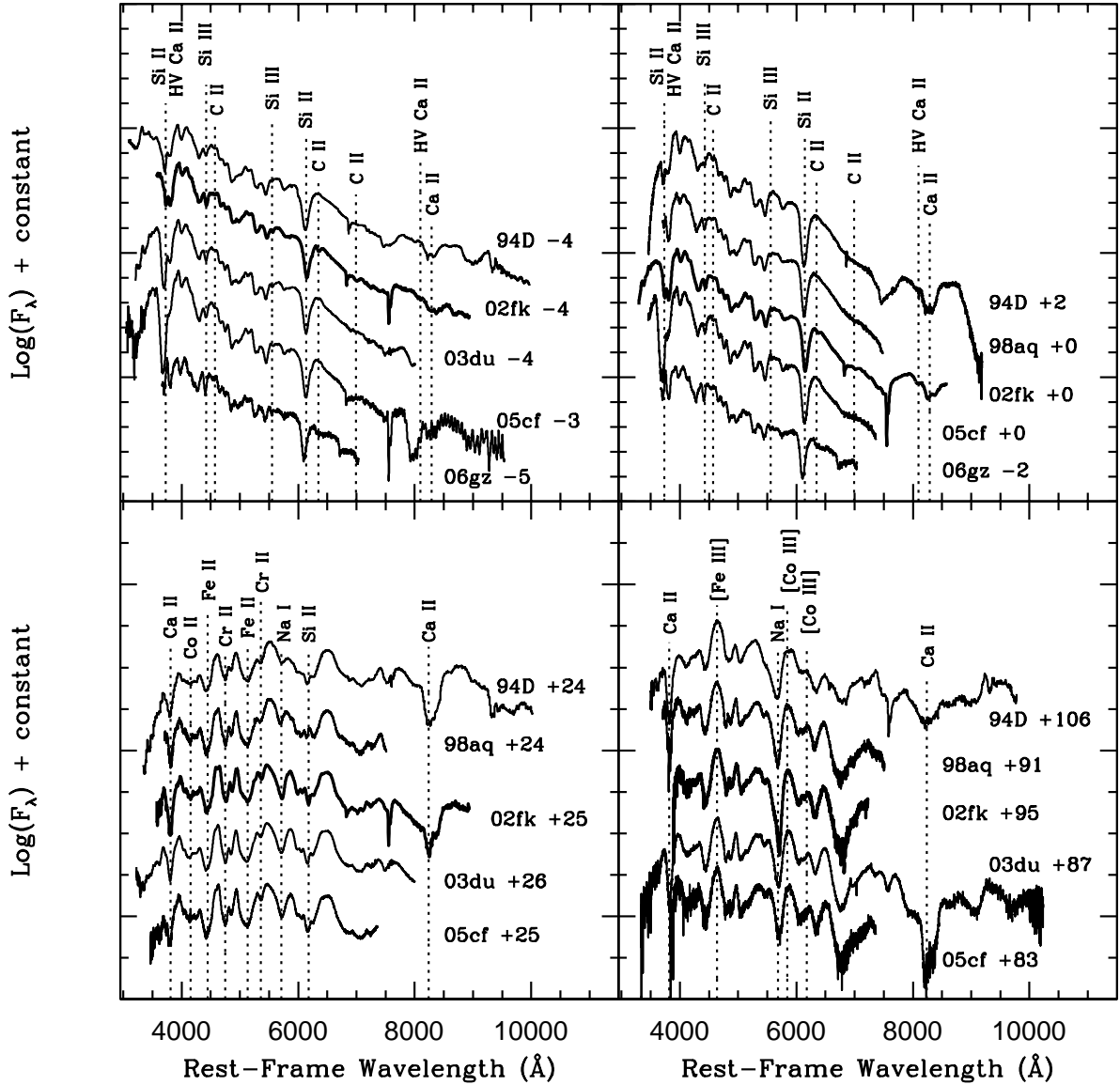}
\caption{Spectrum of SN~2002fk at -4, 0, +25 and +95 days after B maximum. We plot for comparison SN~1994D, SN~1998aq, SN~2003du, SN~2006gz,
and SN~2005cf at similar epochs.}
\label{CompSpec_fig}
\end{figure}

In Figure \ref{CompSpec_fig} we compare the spectra of SN~2002fk with three well-observed SNe with similar $\Delta m 15$:
SN~1998aq \citep{branch03}, SN~2003du \citep{stanishev07}, and SN~2005cf \citep{wang09,bufano09}, the prototypical Branch-normal
SN~1994D \citep{patat96}, and the super--Chandra candidate SN~2006gz \citep{hicken07}. We choose -4, 0, +25 and +95 days after
maximum light as comparison epochs. All the spectra are in the host galaxy rest-frame and were reddening corrected as described 
in Section \ref{opcolors_sec}.

In the top-left of Figure \ref{CompSpec_fig} we show the spectra of SN~2002fk at $\sim-4$ days. The spectrum is characterized by
lines of singly ionized intermediate-mass elements (Si, S, Mg, and Ca), higher-ionized lines of Si~III, C~II, and a strong 
absorption at $\sim3750$ \AA~ that could correspond either to high-velocity Ca~II H\&K, Si~II $\lambda 3858$, or a blend of
both (see Section \ref{models_sec}). At this epoch the spectrum of SN~2002fk is similar to SN~2003du with both SNe showing
strong Si~III absorption lines. Also the super--Chandra candidate SN~2006gz shows prominent Si~III lines. Like SN~2002fk,
SN~2006gz also shows strong C~II $\lambda 6580$ absorption, which is broader than the ones in SN~2002fk. If the HV Ca~II
lines are present in SN~2002fk, they are slightly stronger than the ones observed in SN~1994D, but not as remarkable as the
HV Ca~II lines present in SN~2005cf at this epoch.

In the top-right panel of Figure \ref{CompSpec_fig} we show the spectrum of SN~2002fk near maximum light. At this epoch SN~2002fk
and SN~1998aq share similar spectral shape and features and both SNe show Si~III lines comparable to the lines present in SN~2006gz.
Additionally, SN~2002fk shows normal Ca~II, and maybe weak HV Ca~II lines. In contrast SN~2005cf shows strong HV Ca~II lines. Remarkably,
SN~2002fk still shows clear C~II $\lambda 6580$ and $\lambda 7234$ absorption lines.

In the bottom-left of Figure \ref{CompSpec_fig} we show the spectrum of SN~2002fk at $\simeq +25$ days since maximum. This epoch
corresponds to the beginning of the transition from the photospheric to the nebular phase and the spectra are becoming dominated
by absorption of iron--peak elements (Fe II, Co II, Cr II, etc). The similarity of all the spectra at this epoch is impressive.
However, it is interesting to mention that the Na I absorption in SN~2002fk is slightly stronger than in the others. In the
bottom-right of Figure \ref{CompSpec_fig} we show the spectrum of SN~2002fk and other SNe at $\simeq +95$ days since maximum. At
this epoch the transition from photospheric to nebular phase is almost complete, and all the spectra are very similar showing
forbidden emission lines of iron elements (i.e. [Fe III], [Co III]). SN~2002fk is particularly similar to SN~1998aq, and still
has a slightly stronger Na~I absorption line.

\subsection{Photospheric Expansion Velocity}
\label{photvelocity_sec}

\begin{figure}
\plotone{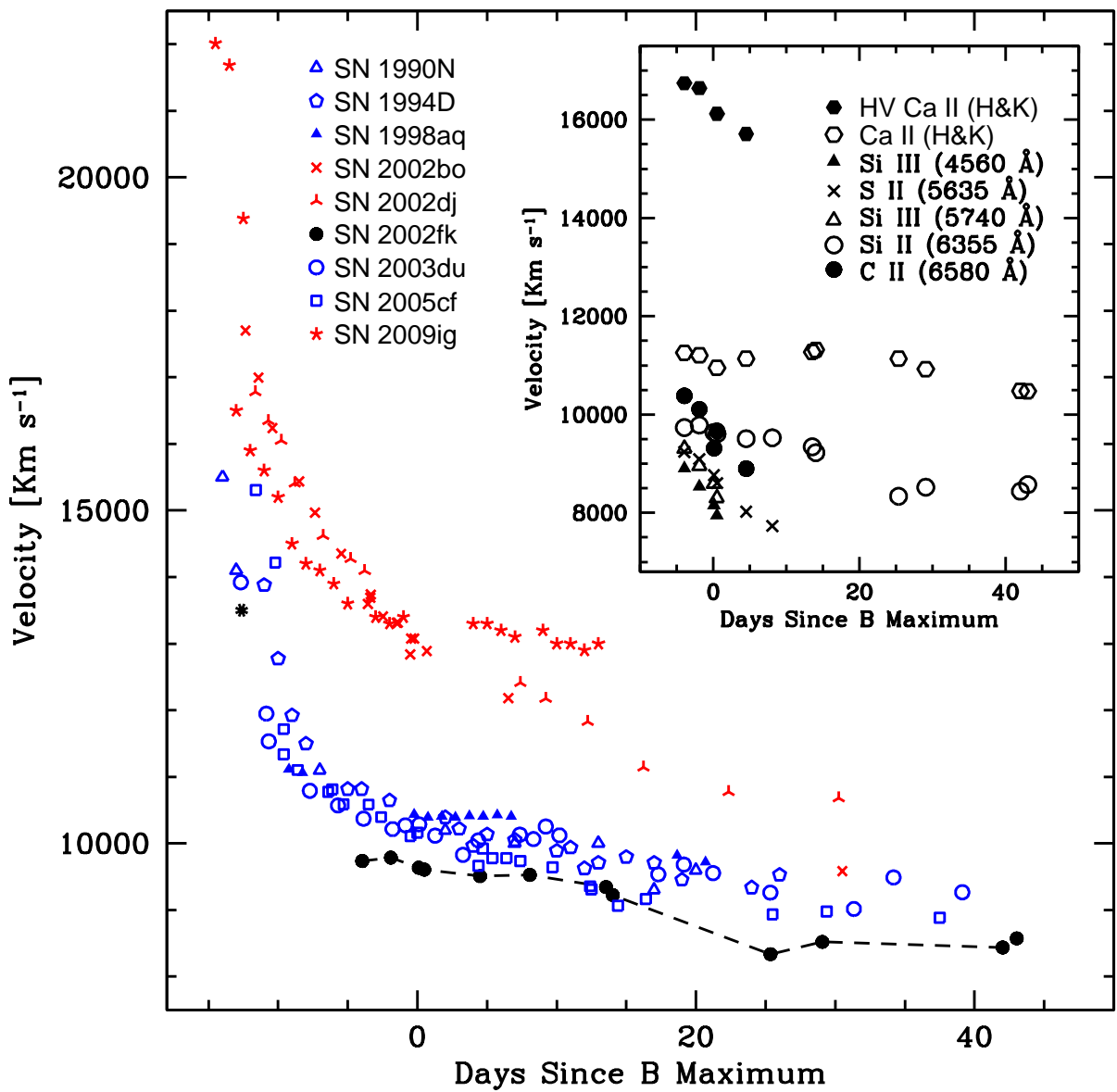}
\caption{Time evolution of the photospheric expansion velocity derived from Si II $\lambda 6355$. Red simbols correspond
to HVG SNe and blue simbols to LVG SNe. Black dots correspond to SN~2002fk, the black asterisc is the velocity of Si II
$\lambda 16~910$ obtained by \citet{marion03} from a near-IR spectrum taken 12 days before maximum. In the inset we show
the time evolution of other optical lines in SN~2002fk.}
\label{SiII_fig}
\end{figure}

In Figure \ref{SiII_fig} we show the photospheric expansion velocity of SN~2002fk measured from the minimum of the Si~II
$\lambda 6355$ line, and a comparison with other SNe~Ia. The Si~II $\lambda 6355$ line is thought to trace the position
of the photosphere \citep{benetti05}. In red we plot three HVG SNe: SN~2002bo \citep{benetti04}, SN~2002dj \citep{pignata08}
and SN~2009ig \citep{foley12a, marion13}. In blue we show the five best observed LVG SNe: SN~1990N \citep{leibundgut91},
SN~1994D \citep{patat96}, SN~1998aq \citep{branch03}, SN~2003du \citep{stanishev07} and SN~2005cf \citep{wang09}. For SN~2002fk
we measure a velocity gradient of $\dot v = 26.3$ km~s$^{-1}$ day$^{-1}$ between $-4$ and $+14$ days since maximum, which falls
in the low end of the distribution of SNe Ia \citep{benetti05}. We also measure the velocity ten days after maximum $v_{10}$ $\sim$
9,400~km~s$^{-1}$. Compared to the \citet{benetti05} sample, SN~2002fk falls ~$3 \sigma$ below the mean (\textless$v_{10}$\textgreater
$\sim$10,300~km~s$^{-1}$ $\pm$ 300~km~s$^{-1}$). From $\dot v$ and $v_{10}$ we unambiguously classify SN~2002fk as a LVG SN. Note
that in the sample of near-IR spectra of SNe Ia presented by \citet{marion03}, SN~2002fk also stands out for its anomalously low
velocity measured from the Mg~II $\lambda~ 10910$ absorption.

In the inset of Figure \ref{SiII_fig} we show velocities measured from other prominent lines in the spectrum. Compared to Si~II
$\lambda 6355$, the Si~III $\lambda 4560$, S~II $\lambda 5635$, and Si~III $\lambda 5740$ lines show lower velocities and
higher velocity gradients, implying that they form below and recede faster into the SN ejecta. We observe that Ca~II $H\&K$
possibly has two components (see Section \ref{models_sec} for a detailed discussion). The main photospheric absorption is
located at $v\sim$ 11,150~km~s$^{-1}$ and shows a flat evolution compared to Si~II $\lambda 6355$. It is possible that a detached
high--velocity Ca~II component is observed four days before maximum with $v\sim$ 16,700~km~s$^{-1}$, and could be observed on
day +5 with $v\sim$ 15,700~km~s$^{-1}$ (see Figure \ref{spectra_fig}). This is similar to the velocity evolution of the Ca~II lines
in HVG SNe.

Before maximum light the minimum of the C~II $\lambda 6580$ absorption line is formed at higher velocites than Si~II $\lambda 6355$
(i.e. at larger radii in the SN ejecta), but near maximum light and thereafter the C~II absorption minimum is measured deeper in the
ejecta than Si~II. This means that there is a significant amount of unburned C~II at the same position, in velocity space, where the
intermediate mass elements (IME) are located. The absorption minimum of C~II $\lambda 6580$ goes deeper in the ejecta with time,
which could be due to emission coming from the red side of the Si~II $\lambda 6355$ line \citep[see][]{fola12}. Although the velocity
estimation from the minimum of the C~II lines has larger associated uncertainties owing to the weakness of the lines, the fact that
both the C~II $\lambda 6580$ and $\lambda 7234$ lines have similar velocities suggests that our measurements are correct. Previous
studies of \citet{parrent11} and \citet{fola12} have shown that the C~II $\lambda 6580$ line is found at higher velocities than the
Si~II $\lambda 6355$ line, and the velocity gradient of these two lines is roughly the same (i.e. they are parallel in velocity versus
time since maximum), which is different from what we observe in SN~2002fk.

\subsection{Presence of Unburned Material}

Nearly 30\% of all SNe Ia show evidence of C~II $\lambda 6580$ before maximum \citep{parrent11,thomas11b,fola12,silverman12b}. In the
middle panels of Figures \ref{models1_fig} and \ref{models2_fig} we present a series of spectra of SN~2002fk from $-3.96$ to $+7.70$
days since $B$--band maximum. The x--axis shows the expansion velocity measured with respect to C II $\lambda 6580$. The presence of
C~II is confirmed by the simultaneous detection of C~II $\lambda 6580$ and $\lambda 7234$ lines at the same expansion velocities
($\sim$ 10,000~km~s$^{-1}$). The detection of both lines at several epochs, together with a possible detection of C~II $\lambda 4745$
in our first two epochs with a similar velocity, gives us confidence in identifying these features as C~II.

Both C~II $\lambda 6580$ and $\lambda 7234$ become weaker and shift to the red with time. Our latest detection of C~II $\lambda 7234$
is at $+4.50$ days past B maximum while the C~II $\lambda 6580$ absorption is detectable even at $+8.07$ days. Notably, this is the
latest detection of C~II in a normal SN Ia and, as discussed in Section \ref{photvelocity_sec}, it is clear indication of the presence
of unburned material mixed with IME's if a homologous expansion of the ejecta is assumed.

\subsection{Spectral Modeling}
\label{models_sec}

We used SYN++ \citep{thomas11a}, an updated version of SYNOW \citep{fisher97} written in C++, to model the spectra. We modeled
our data, CfA, and BSNIP spectra from $-3.96$ to $+7.70$ days since maximum.

In SYN++, one computes a spectrum by specifying the location and optical depth for a given set of ions. The input parameters for
SYN++ are the photospheric velocity, the optical depth, the e-folding length of the opacity profile, the maximum and minimum
cut--off velocity for each ion, and the Boltzmann excitation temperature for parameterizing line strengths.

\begin{figure}
\plotone{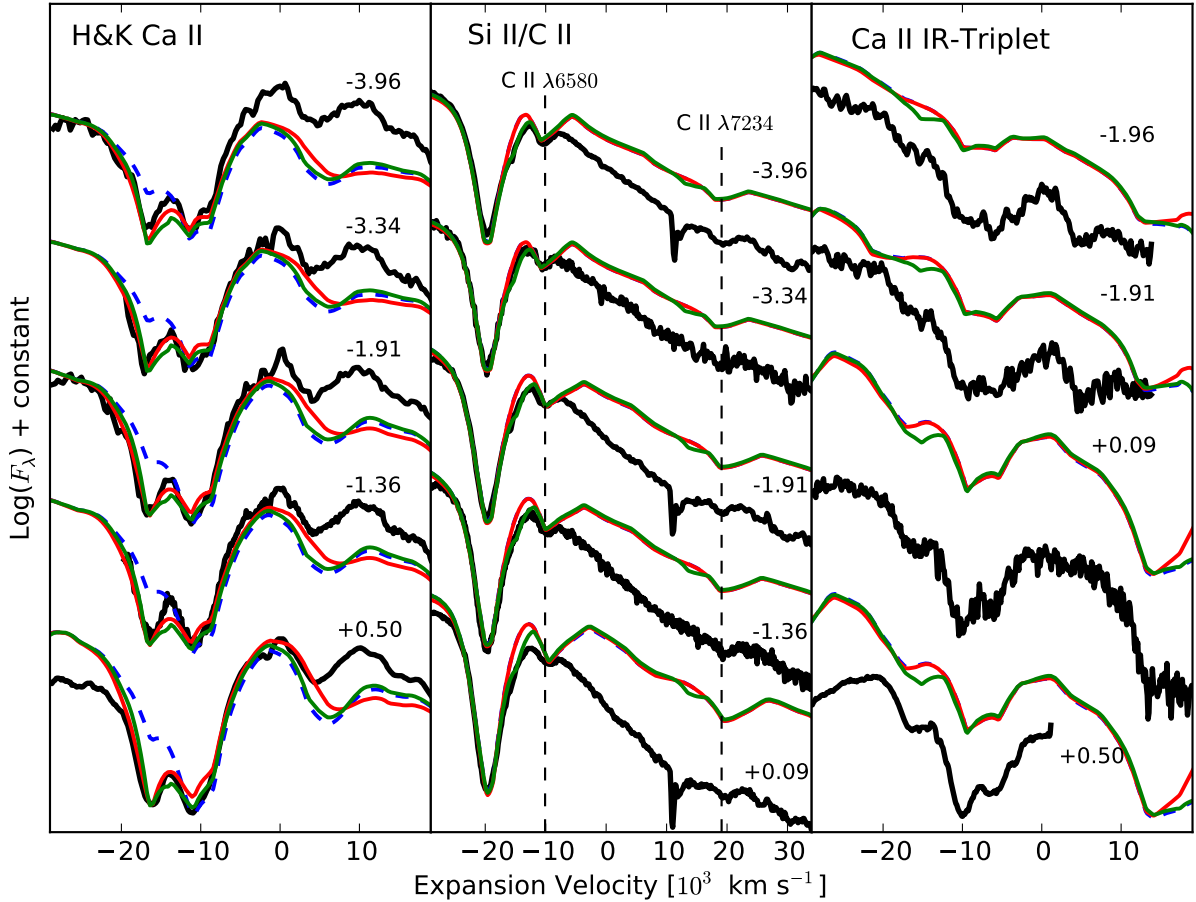}
\caption{SYN++ models of the SN~2002fk spectra. Data are shown in black. Left panel: The blue--dashed lines correspond
to models of $H\&K$ Ca II lines using $T_{exc}$ of Si II $\sim 12,500~K$, which is close to $T_{phot}$, with no HV Ca II
component. In red we show models using $T_{exc}$ of Si II $\sim7,000~K$, and in green we show models using $T_{exc}$
of Si II $\sim 12,500~K$, which is close to $T_{phot}$, and including a HV Ca II component. The x--axis is the expansion
velocity measured with respect to the $H\&K$ Ca II line. Middle panel: in red we show models using $T_{exc}$ of Si II
$\sim 7,000~K$. In green we show models using $T_{exc}$ of Si II $\sim 12,500~K$, which is close to $T_{phot}$, and
including a HV C II component to get a better fit to the red side of the Si II line. The x--axis is the velocity measured
with respect to C~II $\lambda 6580$. Right panel: as in the left panel but for the Ca II IR--triplet. The x--axis is the
expansion velocity measured with respect to the Ca II triplet. The phase of each spectrum is on the right side of each panel.}
\label{models1_fig}
\end{figure}

\begin{figure}
\plotone{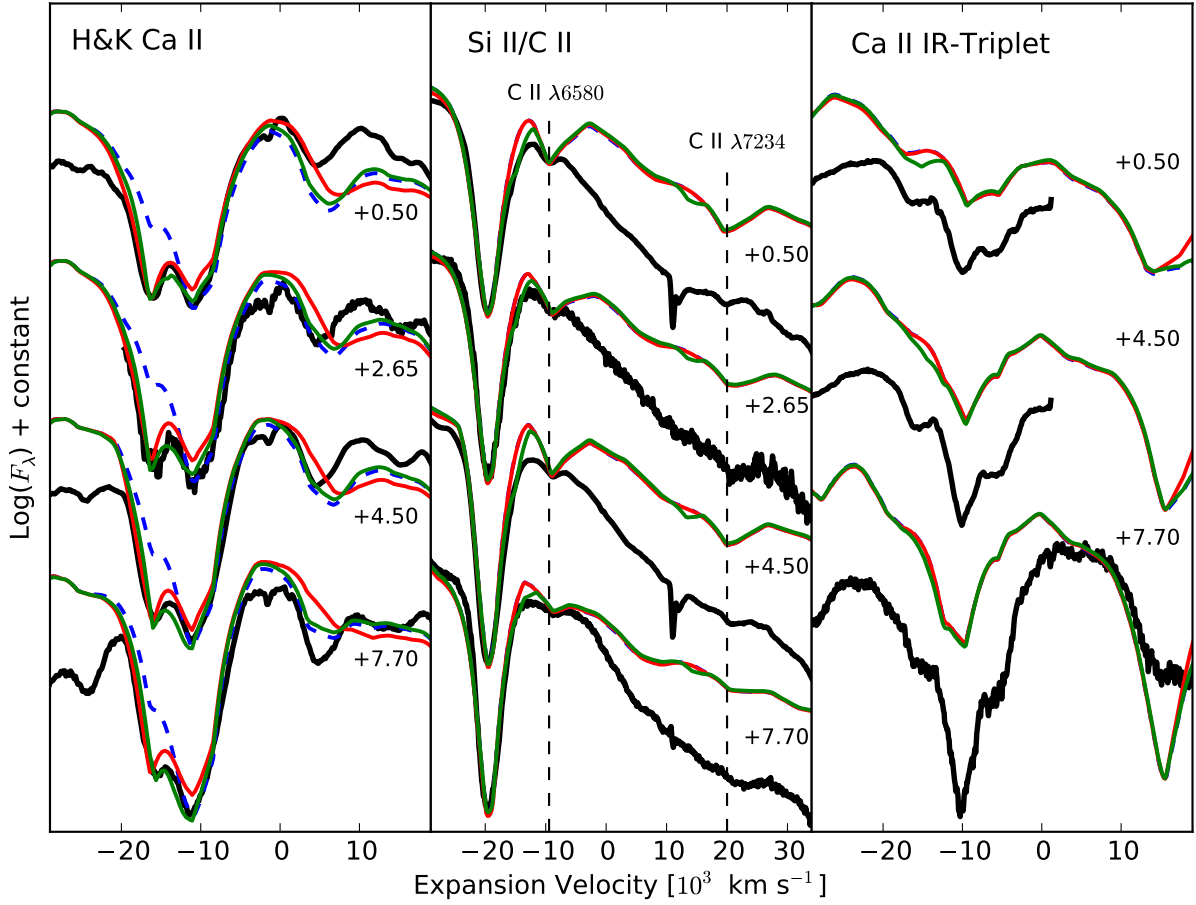}
\caption{SYN++ models of the SN~2002fk spectra. In all three panels we show the same models as in Figure \ref{models1_fig}
but for later phases. The phase of each spectrum is on the right side of each panel.}
\label{models2_fig}
\end{figure}

\begin{figure}
\plotone{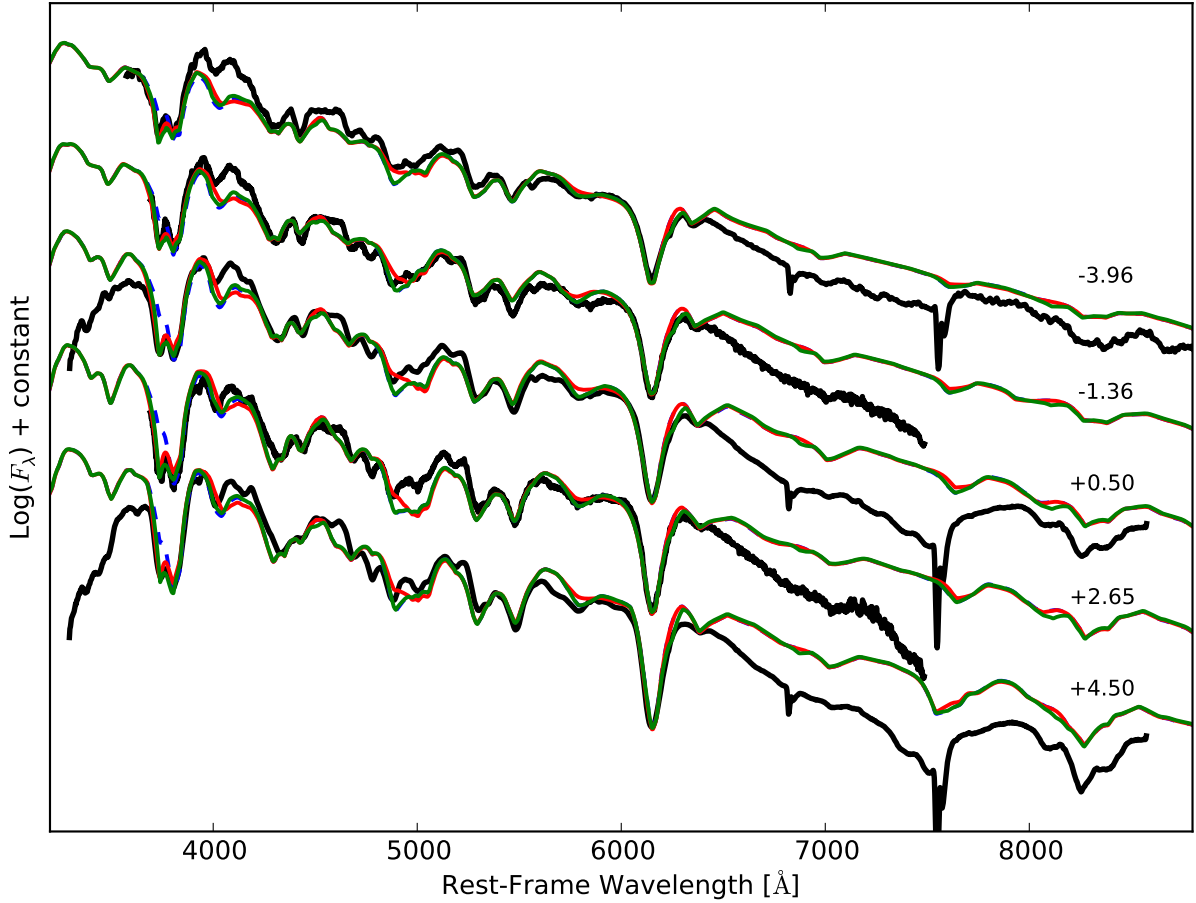}
\caption{SYN++ models of the observed spectra of SN~2002fk spectra, shown in black. The blue--dashed lines correspond to
models using $T_{exc}$ of Si II $\sim 12,500~K$ with no HV components. In red we show models using $T_{exc}$ of Si II
$\sim7,000~K$ with no HV components, and in green we show models using $T_{exc}$ of Si II $\sim 12,500~K$, and including
HV components of Ca~II, and C~II. These are the same models of Figure \ref{models1_fig}, now showing the full spectral
range for representative phases. The phase of each spectrum is on the right side.}
\label{models3_fig}
\end{figure}

Recently \citet{foley13} presented convincing evidence, though not definitive, that the absorption on the blue side of the $H\&K$
Ca II line could be produced by Si II $\lambda 3858$, challenging the traditional interpretation that this absorption is caused
by a HV Ca II line. Here we attempt to test both hypotheses. To do this we fit a SYN++ model to all the spectra from $-3.96$ to $+7.70$
days since $B$--band maximum. In Figure \ref{models3_fig} we present some examples of our models which overall provide good fits to the
observed spectra. We began our modeling by fitting only the photospheric lines, especially Si~II and Ca~II. The characteristic blackbody
temperature ($T_{phot}$) in our models was $\sim 12,800~K$, and we set the excitation temperature ($T_{exc}$) of most of the ions at
$500~K$ below $T_{phot}$. For the high ionizaton lines like Si~III and Fe~III we set $T_{exc}$ to be the same as $T_{phot}$. Then,
in order to get a good fit to the absorption on the blue side of the $H\&K$ line without appealing to a HV Ca~II line, we had to modify
$T_{exc}$ for Si~II to $\sim7000~K$. We note that small variations around this value can change dramatically the strength of Si II
$\lambda 3858$ but not the strength of Si II $\lambda 6355$. Alternatively, we also modeled this feature by adding a HV Ca II component
with $T_{exc}$ set equal to the rest of the ions without modifying the rest of the parameters. Both approaches produce equally good models
(see Figures \ref{models1_fig}, \ref{models2_fig}, and \ref{models3_fig}).

The \citet{foley13} approach requires fewer parameters and reproduces the observations equally well. Remarkably, the velocity and optical
depth of Si II $\lambda 6355$ proved quite insensitive to $T_{exc}$. However, the low $T_{exc}$ required to fit the Si~II $\lambda 3858$
line seems slighty odd for a relatively bright SN with low {\it R}(Si~II) (i.e. hot ejecta) and high ionization lines like Si III and Fe III.
Additionally, there is strong evidence that supports the presence of HV Ca~II in ``normal'' SNe Ia
\citep{kasen03,gerardy04,mazzali05,stanishev07,tanaka08,childress13}. These features are clearly observed at very early phases ($\sim -7$ days)
in both Ca II $H\&K$ and the Ca II IR--triplet, but near maximum they are not always easy to identify \citep{mazzali05}. We conclude that,
while both Si II $\lambda 3858$ and HV Ca II produce equally good fits to the data within the SYN++ framework, more realistic radiative
transfer models will be necessary to distinguish which line is responsible for the absorption on the blue side of the $H\&K$ line.

As can be seen in Figures \ref{models1_fig} and \ref{models2_fig}, the red side of the Si II $\lambda 6355$ profile is not properly
fit by Si II alone (shown in red). To obtain a better fit we added a C~II HV component (shown in green). In the case of our spectra,
which are not corrected for telluric absorption, this feature on the red side of the Si II profile could be due to the lack of such
corrections. However, the CfA spectra obtained on days -3.34 and -1.36, which do include corrections for telluric lines, are best
modelled with a HV C~II component. Interestingly, the HV C~II is at nearly the same velocity of the HV Ca~II ($\sim$ 16,000~km~s$^{-1}$).
The presence of a HV C~II was previously suggested by \citet{fisher97}, and is not completely unexpected since unburned carbon material
should be present in the outer layers of SNe Ia. We flag the presence of HV C II as possible but not as certain.

\section{DETERMINATION OF THE HUBBLE CONSTANT USING THE CEPHEID DISTANCE TO NGC~1309}
\label{h0_sec}

With the relative Cepheid distance between NGC 1309 and NGC 4258 of \citet{riess11}
($\mu_{NGC~1309}-\mu_{NGC~4258}=3.276 \pm 0.05$), the maser distance to NGC 4258 of \citet{humphreys13}
($\mu_{NGC~4258}=29.40 \pm 0.06$),
the peak brightness and reddening of SN~2002fk derived in Section \ref{reddening}, it is straightforward to 
calculate the Hubble constant ($H_0$) via the formula 
$H_{0} = 10^{0.2(M_{max}^{1.1}+25-\alpha)}$, where $M_{max}^{1.1}$ is the absolute magnitude of SN~2002fk corrected for
reddening and $\Delta m 15$, and $\alpha$ is the corresponding zero point of the Hubble diagram. In the optical ($BVI$),
we adopt the decline rate versus luminosity ($\Delta m 15$/$M_{max}$) relations and zero points from \citet{phillips99}, 
who employ the same prescription
that we use here for luminosity and reddening corrections. In the near-IR ($JHK$) we adopt $\Delta m 15$/$M_{max}$ relations
and zero points derived from two independent analysis of nearly the same near-IR CSP observations,
one by \citet{fola10} and the other by \citet{kattner12}.

In Table \ref{H0_tab} we summarize all the adopted parameters and the resulting values for $H_{0}$.
The statistical uncertainty in $H_{0}$ was computed from a Monte Carlo simulation. Our code computes $10,000$ simulated values
of $H_{0}$ assuming a Gaussian distribution in peak magnitude, reddening, $\Delta m15$, the slope of the 
$\Delta m 15$/$M_{max}$ relationship, and the zero point of the Hubble diagram.
The systematic error includes 0.08 mag in distance modulus to NGC 1309, 0.03 mag in the photometric zero point, and
the intrinsic dispersion in the $\Delta m 15$/$M_{max}$ relationship from \citet{phillips99}, \citet{fola10},
and \citet{kattner12}.


As can be seen in Table \ref{H0_tab}, the $H_0$ values that we derive from the optical bands ($BVI$) are very consistent 
with each other, allowing us to combine them into an $BVI$ average of $H_{0}(BVI)=69.6\pm{2.1}~(\pm 5.0$ systematic) \kmsMpc,
where the systematic error includes 0.08 mag in distance modulus to NGC 1309, 0.03 mag in the photometric zero point, 
and the intrinsic dispersion in the $\Delta m 15$/$M_{max}$ relationship including the variance in each band
and band-to-band covariances.
Our value is 0.8 $\sigma$ lower than the one published by \citet{riess11}, $H_{0}=73.8\pm{2.4}$ \kmsMpc,
from eight SNe Ia with precise Cepheid distances. This is due to the fact that SN~2002fk lies in the bright 
side of their luminosity distribution. In fact, using the values listed in Table 3 of \citet{riess11} and their Eq. (4), 
we find $H_{0}=68.5$ \kmsMpc for this same SN, in very good agreement with our determination. 
Thus, using SN~2002fk alone to determine $H_{0}$ is expected to yield
a lower value than the fit to all eight SNe in their sample.


Table \ref{H0_tab} shows that our near-IR estimate of $H_{0}$ is sensitive to the adopted calibration: 
using \citet{kattner12}, we obtain 
$H_{0}(JH)=66.7\pm{1.0}~(\pm 3.5$ systematic) \kmsMpc, whereas if we employ \citet{fola10} we get
$H_{0}(JHK)=63.0\pm{0.8}~(\pm 2.8$ systematic) \kmsMpc. 
This discrepancy is mainly due to the different zero points obtained from their analysis, 0.17 and 0.07 mag 
in $J$ and $H$, respectively (see Table \ref{H0_tab}), which arise from the different techniques employed to
measure peak magnitudes. While \citet{kattner12} directly measured peak magnitudes based on a cubic-spline
interpolation to seven exquisitely observed SNe with observations starting before maximum, \citet{fola10} included peak magnitudes
extrapolated from a template light curve fit.
As mentioned by \citet{kattner12} the large diversity in early time near-IR light curve shapes 
can lead to significant errors when fitting template light curves to derive peak brightnesses.

Despite the differences in $H_{0}$ caused by the near-IR calibration adopted, it appears that the near-IR
yields somewhat lower (6-9\%) $H_{0}$ values than the optical. How significant is this difference? 
Given that the systematic error for this particular SN is almost fully correlated band to band
(owing to the same Cepheid distance and the fact that the intrinsic dispersion of SNe Ia is 
correlated band to band), the significance can be estimated from the statistical 
uncertainty alone, which yields 1.2 and 2.9 $\sigma$ for \citet{kattner12} and \citet{fola10}, respectively.
It will be essential to further examine this issue by expanding the sample of near-IR SN light curves
with good sampling near maximum, which should lead to a more robust determination of the zero point
of the Hubble diagram and the $\Delta m 15$/$M_{max}$ relationship. Such effort should be rewarded with
a significantly more precise determination of $H_{0}$: as can be seen in Table \ref{H0_tab}, the 
near-IR yields smaller statistical uncertainties due to a smaller sensitivity to extinction, decline rate
correction, and higher precision in the zero point of the Hubble diagram. As opposed to the
optical bands where the zero point is subject to inaccuracies caused by the heterogeneity in the various
photometric systems employed, our near-IR zero point is based in a single
homogeneous Carnegie photometric dataset, an essential ingredient in lowering the
statistical uncertainty in $H_0$.

An independent value of $H_{0}=69.77^{+2.07}_{-2.07}$ was obtained by \citet{sullivan11} by combining the SNLS3 data with the full
WMAP7 power spectrum and the Sloan Digital Sky Survey luminous red galaxy power spectrum (DR7). Similarly, \citet{mehta12} obtained
$H_{0}=69.8 \pm 1.2$ by fitting a $\Lambda CDM$ model to the DR7 and WMAP7 data.
Studies of galaxy clustering from the SDSS--III Data Release 8 (DR8) yield $H_{0}=70.5\pm1.6$ \citep{ho12}. 
More recently, the first results based on the Planck satellite measurements of the CMB temperature and lensing-potential
power spectra find a low value of $H_0=67.3\pm1.2$ \citep{ade14}.
All these experiments
reveal some tension with the $H_0=73.8\pm2.4$ value obtained by \citet{riess11}.
Our analysis provides the first determination of $H_0$ from near-IR SNe Ia data, and suggests that 
the near-IR could help to alleviate this tension.


\section{DISCUSSION AND CONCLUSIONS}
\label{discussion}

We present early and complete optical ($BVRI$) and near-IR ($JHK$) light-curves of SN~2002fk starting 12 days before maximum light
through 122 days past maximum, along with a series of 15 optical spectra from -4 to +95 days since maximum.
Our study reveals the following distinguishing properties of SN~2002fk:

$\bullet$ Low decline rate ($\Delta m 15$=1.02) of the B lightcurve, consistent with a higher than average peak luminosity
($M_B$=-19.49).

$\bullet$ Negligible host galaxy reddening as inferred from the observed colors.


$\bullet$ Slightly bluer than average $B-V$ color near maximum.

$\bullet$ Presence of Si~II 6355 with a relatively low (9,500~\kms) expansion velocity near maximum, and a small velocity gradient,
$\dot v_{Si~II}=26$~km~s$^{-1}$~day$^{-1}$. 

$\bullet$ Blue $U-B$ and a red $V-I$ early-time color, expected for LVG SNe~Ia \citep{cartier11}.

$\bullet$ Presence of C~II lines in the early-time spectra expanding at ~11,000~\kms persisting until ~8 days past maximum light
with a velocity $\sim$9,000~\kms, and possible HV component with 16,000~\kms.

$\bullet$ Possible presence of a HV Ca II $H\&K$ component in the early-time spectra expanding at $\sim 16,000$~\kms and persisting until $\sim$ 8 
days past maximum, with similar velocity of a possible HV C~II component
\citep[an alternative explanation for the HV Ca II is Si II $\lambda 3858$; see][]{foley13}.

In the off-center explosion models of \citet{maeda10c}, SNe~Ia with low Si~II velocity gradients are those with the ignition region oriented
towards the observer. Therefore, we expect a negative (blueshifted) velocity shifts of the [Fe II] $\lambda 7155$ and [Ni II] $\lambda 7378$
nebular lines ($V_{neb}$). \citet{silverman13} have recently shown that this is indeed the case for SN~2002fk ($V_{neb}=-2060$~\kms),
lending support to the \citet{maeda10c} models, the \citet{cartier11} relations (see their Fig. 2), and \citet{2012ApJ...754L..21F} work.
\cite{parrent11} and \cite{fola12} have noted that SNe showing C~II in their spectra are grouped preferentially in the bottom of the Si~II
expansion velocity distribution. Our observations of SN~2002fk are consistent with such findings. The connection between viewing angle of an
off-center explosion and the presence of C~II in the early time spectrum suggests that the observation of C~II could be also due to a
viewing angle effect. A possible explanation could be that when the ignition region is directed towards us, the burning front leaves
some pockets of unburned material in the direction of the observer. Our observations indicate also that SN~2002fk is quite luminous 
and somewhat bluer than normal, which might also be a signature of an ignition region oriented towards the observer.

The possible presence of a HV Ca II component in the early-time spectra expanding at $\sim 16,000$~\kms persisting until $\sim$ 8
days past maximum, at a similar velocity of a possible detection of a HV C~II component, suggests that these HV features
may arise from density enhanced regions containing a mixture of Ca~II and C~II.

Our photometry shows good agreement with the optical magnitudes published by \citet{riess09b}, except in the $I$-band where we find an
offset of 0.07 mag. We do not have an explanation for this worrisome discrepancy except that LOSS employed an $I$ filter with an
extended red tail compared to the standard bandpass. 

Adopting the Cepheid distance to NGC 1309 recently measured by \citet{riess11} and \citet{humphreys13},
we derive optical and near-IR absolute magnitudes for SN~2002fk.
Using the $BVI$ absolute magnitude/decline rate calibration of \citet{phillips99} we solve
for the Hubble constant, obtaining a consistent value of $H_{0}=69.6 \pm 2.1$~($\pm~5.0$ systematic), which is
0.8 $\sigma$ lower than the one published by \citet{riess11}, $H_{0}=73.8 \pm 2.4$ from eight
SNe Ia with precise Cepheid distances. This is due to the fact that SN~2002fk lies on the bright side of their
luminosity distribution. 

We proceed in a similar manner for the $JHK$ absolute magnitudes to derive the first $H_{0}$ 
value based on near-IR measurements of a SN Ia.
As opposed to the optical method that relies on many different
photometric calibrations containing significant systematic differences (as shown in
Appendix \ref{local_standards}), the near-IR has the great advantage 
that it is founded on a homogeneous Carnegie photometric dataset including 
(1) the Persson standards \citep{persson02}, 
(2) the CSP Hubble diagram \citep{fola10,kattner12}, and 
(3) the calibrating CATS SN 2002fk.
Our near-IR $H_{0}$ values rely on two CSP different 
absolute magnitude/decline rate relationships.
Using the \citet{kattner12} relation, we obtain
$H_{0}(JH)=66.7\pm{1.0}~(\pm 3.5$ systematic) \kmsMpc, whereas employing \citet{fola10} we get
$H_{0}(JHK)=63.0\pm{0.8}~(\pm 2.8$ systematic) \kmsMpc.
This discrepancy is mainly due to the different zero points obtained from their analysis, 0.17 and 0.07 mag
in $J$ and $H$, respectively, which arise from the different techniques employed to
measure peak magnitudes. More importantly, it appears that the near-IR yields somewhat
lower (6-9\%) $H_0$ values than the optical, with a significance of 1.2 and 2.9 $\sigma$ depending on 
the near-IR calibration adopted. It is essential to further examine this issue
by (1) expanding the sample of high-quality near-IR light curves of SNe in the Hubble flow, and 
(2) increasing the number of nearby SNe with near-IR SN light curves and 
precise Cepheid distances, which affords the promise to deliver a more precise determination of $H_0$.


\acknowledgments

\noindent
We thank the anonymous referee for comments that helped to improve our work. R.C. acknowledges support by CONICYT through
``Programa Nacional de Becas de Postgrado'' grant D-2108082, and by the Yale-Chile fellowship in astrophysics. R.C., G.P.,
M.H., P.Z., F.F., J.M., A.C. acknowledge support provided by the ``Millennium Center for Supernova Science'' through grant
P10-064-F  and the ``Millennium Institute of Astrophysics (MAS)'' through grant IC120009 of the
``Programa Iniciativa Cient\'ifica Milenio del Ministerio de Econom\'ia, Fomento y Turismo de Chile''. F.F., and G.P.
acknowledge support from FONDECYT through grants 3110042, and 11090421. The authors acknowledge Rolling~C. Thomas for his
support with syn++, and Ryan Foley for kindly providing spectra of SN~2009ig. We also thank Santiago Gonz\'alez-Gait\'an 
and Adam Riess for their useful input and feedback.

The research presented in this article made use of SUSPECT \footnote[14]{See http://suspect.nhn.ou.edu/~suspect/} Online
Supernova Spectrum Archive.

\appendix

APPENDICES

\section{CROSS-CAMERA SUBTRACTIONS}
\label{cam_subtractions}

Since our template images in $UBVRI$ were acquired with a different instrument than the ones used for the SN imaging
(see Table \ref{Phot_tab}), there is a potential systematic error caused by our cross-camera subtractions. To assess this error
we performed the following intrepid test: we subtracted the $V$ template from the $B$ SN images, the $B$ template from the $V$
SN images, the $V$ template from the $R$ SN images, and the $R$ template from the $I$ SN images. Then we measured instrumental
magnitudes with a 3 arcsec aperture both in the crossed subtracted images and those performed with the same filter. The magnitude
differences yielded by this test amounted to 0.001-0.003 mag. Given that these cross subtractions involve wavelength differences
$\sim$1,000 \AA~between templates and SN images, the actual error caused by using templates that differ by $\sim$10 \AA~in central
wavelenghts (as in our case) from the SN images can be ignored. In the near-IR this potential uncertainty does not exist because
the templates images were obtained with the same instrument employed for the SN.

\section {LOCAL STANDARDS \& PHOTOMETRIC ZERO POINTS}
\label{local_standards}


Given that there are other sources of optical photometry available in the literature for SN~2002fk, it is necessary to verify
the compatibility of the individual photometric calibrations before attempting to merge the photometry. Initially we started
this exercise by comparing the magnitudes of the four local standard stars in common (stars 5, 6, 7, 8) between us (LCO/Swope)
and LOSS. To our surprise we discovered very good agreement in BVR but a significant systematic offset $\sim$0.07 mag in the
$I$-band. When we noticed this discrepancy, we proceeded to obtain additional measurements of the local standards with the
CTIO/0.9-m telescope (two nights) and the CTIO/PROMPT telescopes (two nights) under photometric conditions. Given that we
found very good agreement ($\le$0.015 mag) in $BVRI$ among all three calibrations and a reasonable ($\sim$0.06 mag) consistency
in the $U$-band between LCO/Swope and CTIO/0.9--m (no $U$ measurements could be obtained with CTIO/PROMPT) we decided to
combine all of our $UBVRI$ values into the single CATS calibration, which is presented in Table \ref{OpSequence_tab}.

Our internal consistency can be seen in detail in Table \ref{SysDiff_tab} which summarizes the differences in each filter
between the magnitudes of the local standard stars measured with our three instruments (LCO/Swope, CTIO/0.9-m, CTIO/PROMPT)
and our average CATS calibration. We find excellent agreement in $BVRI$ ($\le$0.015 mag). This is remarkable considering
that the CTIO/PROMPT data were reduced with a completely independent pipeline by one of us (GP). In the $U$-band the $\sim$0.06
mag difference between LCO/Swope and CTIO/0.9--m (no $U$ measurements could be obtained with CTIO/PROMPT) is not surprising
given that $U$--band photometry is generally difficult to calibrate as a result of filter mismatches or differences in the
sky transmission (i.e. effective bandpass), and large systematic differences are common.

Having demonstrated our own internal consistency (with the above mentioned caveat in $U$) we return to the issue of comparing
our photometry with the external measurements of the local standards obtained by LOSS \citep{riess09b, ganesha10} and CfA3 
\citep{hicken09}. This comparison is presented in Table \ref{SysDiff_tab} which gives magnitude differences between the local
standard stars measured by CATS, LOSS, and CfA3. Overall we found good agreement between LOSS and CATS ($\le$0.02 mag), with
the notable exception of the $I$-band where the LOSS magnitudes are systematically brighter by $\sim 0.07$ mag. We do not
have an explanation for this worrisome discrepancy except that LOSS employed on the Nickel telescope an $I$ filter with an
extended red tail compared to the standard bandpass [see Fig. 1 in \citet{ganesha10}]. Perhaps this is also the cause of the
systematic differences $\sim$0.1 mag in the $I$--band between LOSS and CfA2 \citep{jha06} reported by \citet{ganesha10} for
some SNe. When we compare CATS and CfA3, we find better agreement in $I$ (CfA3 is between CATS and LOSS). However, what is
more concerning is that CfA3 is consistently brighter than CATS by $\sim 0.03-0.04$ mag in all $UBVRI$ filters. While LOSS
is halfway between CATS and CfA3 in $B$, LOSS is much closer to CATS than CfA3 in $V$ and $R$. We cannot claim which one
is correct, but the advantage of the CATS calibration is that it was obtained with three different cameras, two independent
pipelines, and two reducers.

Based on this analysis, we adopt a systematic uncertainty of $0.03$ mag for our photometric zero points. Unfortunately, we
cannot apply this approach to the near-IR because the 2MASS magnitudes are not sufficiently precise. Hence, we adopt the
same uncertainty of $0.03$ mag.

\section{S-CORRECTIONS}
\label{s_corrections}

SN magnitudes obtained with different instruments can potentally lead to significant systematic errors due to the non-stellar
nature of the SN spectrum. These effects, a.k.a. S-terms, were computed by \citet{stritzinger02} for the normal Type Ia SN~1999ee.
As shown in their Fig. 4, the S-corrections change with time as the SN spectrum evolves. The systematic errors for the CTIO/0.9-m
camera are generally small in $BVRI$, from 0.01 mag near maximum to 0.04 mag at late times ($\sim$60 days past maximum), suggesting
that this instrument provides a good match to the standard system \citep{bessell90}, while for the YALO instrument the systematic
errors are larger ($\sim$0.05-0.08 mag) owing to significant instrumental departures from the standard bandpasses. In order
to quantify this effect in SN~2002fk, we took advantage of the fact that the object was measured with seven different
instruments allowing us to calculate magnitude differences among all of them after removing the systematic differences
caused by the photometric zero-points. This empirical approach showed that the S-terms amount to 0.02 mag at maximum
light and 0.03 mag at later epochs (40-80 days after peak), allowing us to to attach a systematic error to our $BVRI$
photometry. Its is encouraging that this empirical approach yields similar values to those computed by \citet{stritzinger02}
for the CTIO/0.9-m camera. Likewise, in the near-IR we estimate the systematic errors caused by the S-terms to be 0.02 mag at maximum,
from a comparison of the SN magnitudes measured with Baade/Classic-Cam and du Pont/WIRC. We cannot apply this technique at later
times in the near-IR since SN~2002fk was observed with a single instrument, but we can safely adopt the 0.03 uncertainty measured
in the optical. Most likely, this is an upper limit to the systematic errors in the near-IR, considering that the instrument
detector and filters used with WIRC and Classic-Cam were essentially the same as those employed by \citet{persson98} in the
establishment of the standard system.

\clearpage

\begin{deluxetable} {cccccc}
\rotate
\tablecolumns{6}
\tablenum{1}
\tablewidth{0pc}
\tablecaption{Photometric Observations of SN~2002fk.\label{Phot_tab}}

\tablehead{
\colhead{Observatory} &
\colhead{Telescope} &
\colhead{Instrument} &
\colhead{Detector} &
\colhead{Plate Scale} &
\colhead{Filters} \\
\colhead{} &
\colhead{} &
\colhead{} &
\colhead{} &
\colhead{(arcsec pixel$^{-1}$)} &
\colhead{}}
\startdata

LCO   & Swope   & CCD                         & SITe 2048$\times$3150                   & 0.435 & $UBV(RI)_{KC}$ \\
LCO   & du Pont & WFCCD                       & Tek  2048$\times$2048                   & 0.774 & $BV(I)_{KC}$ \\
LCO   & du Pont & CCD                         & Tek  2048$\times$2048                   & 0.259 & $UBV(RI)_{KC}$ \\
LCO   & Baade   & LDSS-2\tablenotemark{a}     & SITe 2048$\times$2048                   & 0.380 & $BV(R)_{KC}$ \\
CTIO  & 0.9m    & CCD                         & 2048$\times$2048                        & 0.396 & $UBV(RI)_{KC}$ \\
CTIO  & 1.5m    & CCD                         & 2048$\times$2048                        & 0.440 & $BV(RI)_{KC}$ \\
CTIO  & PROMPT  & CCD\tablenotemark{b}        & Alta U47 E2V 1024$\times$1024           & 0.600 & $BV(RI)_{KC}$ \\
&&&&& \\
LCO   & du Pont & WIRC\tablenotemark{c}       & HAWAII-1 1024$\times$1024               & 0.196 & $J_SHK_S$ \\
LCO   & Baade   & Classic-Cam                 & NICMOS3  256$\times$256                 & 0.094 & $J_SHK_S$ \\

\enddata
\tablenotetext{a}{\cite{allington-smith94}}
\tablenotetext{b}{\cite{pignata11}}
\tablenotetext{c}{\cite{persson02}}
\end{deluxetable}

\clearpage

\begin{deluxetable} {ccccccc}
\tablecolumns{7}
\tablenum{2}
\tablewidth{0pc}
\tablecaption{Spectroscopic Observations of SN~2002fk.\label{Spec_tab}}
\tablehead{
\colhead{Date(UT)} &
\colhead{JD} &
\colhead{Phase} &
\colhead{Instrument/} &
\colhead{Wavelength}&
\colhead{Resolution} &
\colhead{Exposure} \\
\colhead{} &
\colhead{-2,400,000} &
\colhead{(days)} &
\colhead{Telescope} &
\colhead{Range (\AA)} &
\colhead{(\AA)} &
\colhead{(s)}}
\startdata

2002 Sep 26 & 52543.84 & -3.96 & LDSS-2/Baade       & 3600 - 9000 & 14 & 120 \\
2002 Sep 28 & 52545.89 & -1.91 & LDSS-2/Baade       & 3600 - 9000 & 14 & 200 \\
2002 Sep 30 & 52547.89 & 0.09  & WFCCD/du Pont      & 3800 - 9200 & 8  & 200 \\
2002 Oct 01 & 52548.31 & 0.50  & FORS1-PMOS/VLT     & 3330 - 8500 & 12 & 2880 \\
2002 Oct 05 & 52552.32 & 4.50  & FORS1-PMOS/VLT     & 3330 - 8500 & 12 & 2880 \\
2002 Oct 08 & 52555.87 & 8.07  & WFCCD/du Pont      & 3800 - 9200 & 8  & 200 \\
2002 Oct 14 & 52561.36 & 13.54 & FORS1-PMOS/VLT     & 3330 - 8500 & 12 & 2400 \\
2002 Oct 14 & 52561.83 & 14.03 & LDSS-2/Baade       & 3600 - 9000 & 14 & 60 \\
2002 Oct 25 & 52573.15 & 25.35 & LDSS-2/Baade       & 3600 - 9000 & 14 & 120 \\
2002 Oct 29 & 52576.87 & 29.07 & LDSS-2/Baade       & 3600 - 9000 & 14 & 90 \\
2002 Nov 11 & 52589.82 & 42.02 & B\&C Spec./Baade   & 3200 - 9200 & 7 & 300 \\
2002 Nov 12 & 52590.83 & 43.03 & B\&C Spec./Baade    & 3200 - 9200 & 7 & 200 \\
2002 Nov 28 & 52606.75 & 58.95 & B\&C Spec./Baade    & 3200 - 9200 & 7 & 200 \\
2002 Dec 02 & 52610.70 & 62.90 & WFCCD/du Pont      & 3800 - 9200 & 8 & 450 \\
2003 Jan 03 & 52642.73 & 94.93 & Mod. Spec./du Pont & 3790 - 7270 & 7 & 450 \\

\enddata
\end{deluxetable}

\clearpage

\begin{deluxetable} {ccccccccccccc}
\rotate
\tablecolumns{13}
\tablenum{3}
\tablewidth{0pc}
\tablecaption{$UBRVI$ Photometric Sequence Around SN~2002fk.\label{OpSequence_tab}}
\tablehead{
\colhead{Star} &
\colhead{$U$} &
\colhead{$N$} &
\colhead{$B$} &
\colhead{$N$} &
\colhead{$V$} &
\colhead{$N$} &
\colhead{$R$} &
\colhead{$N$} &
\colhead{$I$} &
\colhead{$N$} &
\colhead{LOSS\tablenotemark{a}} &
\colhead{CfA3\tablenotemark{b}} \\
\colhead{} &
\colhead{} &
\colhead{} &
\colhead{} &
\colhead{} &
\colhead{} &
\colhead{} &
\colhead{} &
\colhead{} &
\colhead{} &
\colhead{} &
\colhead{name} &
\colhead{name}
}
\startdata

c1& 13.604(0.018) & 4 & 13.424(0.016) & 4 & 12.705(0.014) & 4 & 12.322(0.015) & 4 & 11.951(0.014) & 4 & - & - \\
c2& 14.337(0.020) & 4 & 13.386(0.016) & 4 & 12.346(0.014) & 4 & 11.752(0.015) & 4 & 11.279(0.014) & 4 & - & - \\
c3& 13.248(0.013) & 4 & 13.278(0.025) & 4 & 12.718(0.014) & 4 & 12.366(0.016) & 4 & 12.065(0.014) & 4 & - & - \\
c4& 14.313(0.014) & 4 & 14.230(0.019) & 4 & 13.620(0.013) & 4 & 13.281(0.008) & 4 & 12.961(0.008) & 4 & - & - \\
c5& 16.197(0.009) & 6 & 16.330(0.013) & 8 & 15.771(0.006) & 8 & 15.424(0.007) & 8 & 15.077(0.005) & 8 & 4 & 4 \\
c6& 16.419(0.010) & 6 & 16.437(0.013) & 8 & 15.820(0.006) & 8 & 15.447(0.006) & 8 & 15.080(0.006) & 8 & 2 & 5 \\
c7& 16.209(0.011) & 6 & 16.276(0.008) & 8 & 15.759(0.005) & 8 & 15.429(0.010) & 8 & 15.102(0.006) & 8 & 1 & 6 \\
c8& 17.519(0.018) & 6 & 17.555(0.013) & 8 & 16.948(0.008) & 8 & 16.595(0.007) & 8 & 16.237(0.008) & 8 & 3 & 3 \\
c9&  \nodata      & 4 & 19.083(0.028) & 4 & 17.842(0.021) & 4 & 17.069(0.010) & 4 & 16.425(0.007) & 4 & - & - \\
c10& \nodata      & 4 & 16.329(0.016) & 4 & 15.780(0.014) & 4 & 15.425(0.015) & 4 & 15.122(0.010) & 4 & - & - \\
c11& \nodata      & 4 & 15.221(0.016) & 4 & 14.669(0.014) & 4 & 14.310(0.015) & 4 & 14.013(0.010) & 4 & - & - \\

\enddata
\tablecomments{Errors given in parenthesis, are 1$\sigma$ statistical uncertainties. For each star we indicate N, the number of photometric nights used to calibrate them.}
\tablenotetext{a} {\citet{ganesha10}}
\tablenotetext{b} {\citet{hicken09}}

\end{deluxetable}

\clearpage

\begin{deluxetable} {cccccc}
\tablecolumns{6}
\tablenum{4}
\tablewidth{0pc}
\tablecaption{Color Terms for the five Optical Cameras \label{coefficients_tab}}
\tablehead{
\colhead{Telescope} &
\colhead{$U$} &
\colhead{$B$} &
\colhead{$V$} &
\colhead{$R$} &
\colhead{$I$}
}

\startdata

Swope/CCD          & +0.177  & +0.053  & -0.054 & +0.021  & +0.052  \\
du Pont/WFCCD      & \nodata & +0.125  & -0.045 & \nodata & +0.010  \\
Clay/LDDS-2        & \nodata & +0.132  & +0.046 & \nodata & \nodata \\
CTIO 0.9-m/CCD     & +0.126  & -0.086  & +0.011 & +0.004  & +0.007  \\
CTIO 1.5-m/CCD     & +0.139  & -0.080  & +0.030 & +0.016  & +0.017  \\

\enddata
\tablecomments{Color terms are defined in equations \ref{U_eq}-\ref{I_eq}.}
\end{deluxetable}

\clearpage

\begin{deluxetable} {cccccccc}
\rotate
\tablecolumns{8}
\tablenum{5}
\tablewidth{0pc}
\tablecaption{$UBVRI$ Photometry of SN~2002fk.\label{OpPhotometry_tab}}
\tablehead{
\colhead{Date(UT)} &
\colhead{JD-2,400,000} &
\colhead{$U$} &
\colhead{$B$} &
\colhead{$V$} &
\colhead{$R$} &
\colhead{$I$} &
\colhead{Telescope} }
\startdata

2002 Sep 21 & 52538.8 & 13.523(0.013) & 14.128(0.016) & 14.234(0.014) & 14.149(0.015) & 14.119(0.010) &  Swope \\
2002 Sep 26 & 52543.9 & \nodata       & 13.443(0.012) & 13.515(0.014) & \nodata	      & \nodata	      &  Baade  \\
2002 Sep 30 & 52547.9 & \nodata       & 13.319(0.047) & 13.358(0.019) & \nodata	      & 13.609(0.051) &  du Pont\\
2002 Oct 01 & 52548.9 & 12.792(0.018) & 13.287(0.016) & 13.350(0.014) & 13.335(0.011) & 13.595(0.014) &  Swope  \\
2002 Oct 08 & 52555.9 & \nodata       & 13.675(0.076) & 13.623(0.085) & \nodata	      & 14.003(0.074) &  du Pont\\
2002 Oct 14 & 52561.8 & \nodata       & 14.261(0.017) & 13.904(0.014) & \nodata	      & \nodata	      &  Baade  \\
2002 Oct 20 & 52567.8 & 15.044(0.010) & 14.883(0.011) & 14.239(0.010) & 14.133(0.011) & 14.181(0.010) &  Swope  \\
2002 Oct 21 & 52568.8 & 15.195(0.024) & 15.014(0.011) & 14.355(0.023) & 14.139(0.013) & 14.218(0.023) &  Swope  \\
2002 Oct 24 & 52571.9 & 15.544(0.017) & 15.294(0.014) & 14.409(0.015) & 14.196(0.015) & 14.198(0.015) &  CTIO 0.9-m\\
2002 Oct 25 & 52572.8 & 15.631(0.018) & 15.394(0.016) & 14.500(0.014) & 14.205(0.015) & 14.121(0.014) &  Swope\\
2002 Oct 29 & 52572.9 & \nodata       & \nodata       & 14.511(0.014) & \nodata	      & \nodata	      &  Baade\\
2002 Oct 30 & 52576.9 & \nodata       & 15.782(0.018) & 14.714(0.010) & 14.376(0.014) & \nodata	      &  Baade \\
2002 Oct 08 & 52577.9 & \nodata       & 15.825(0.017) & 14.778(0.019) & \nodata       & 14.123(0.017) &  du Pont\\
2002 Nov 07 & 52585.8 & 16.388(0.018) & 16.238(0.016) & 15.229(0.014) & 14.888(0.015) & 14.539(0.015) &  Swope\\
2002 Nov 08 & 52586.9 & \nodata       & 16.277(0.016) & 15.256(0.014) & 14.939(0.015) & 14.570(0.019) &  Swope \\
2002 Nov 12 & 52590.8 & 16.531(0.020) & 16.396(0.016) & 15.440(0.014) & 15.129(0.015) & 14.840(0.014) &  Swope\\
2002 Nov 18 & 52596.7 & 16.668(0.019) & 16.546(0.016) & 15.640(0.014) & 15.358(0.015) & 15.161(0.014) &  Swope \\
2002 Dec 02 & 52610.7 & \nodata       & 16.819(0.062) & 16.075(0.033) & \nodata	      & 15.785(0.019) &  du Pont \\
2002 Dec 04 & 52612.8 & 17.045(0.023) & 16.830(0.016) & 16.091(0.014) & 15.922(0.015) & 15.887(0.014) &  Swope \\
2002 Dec 07 & 52615.7 & 17.043(0.018) & 16.848(0.016) & 16.128(0.014) & 15.981(0.015) & 16.000(0.014) &  Swope \\
2003 Jan 07 & 52646.7 & 17.803(0.024) & 17.324(0.016) & 16.907(0.010) & 16.954(0.015) & 17.184(0.014) &  Swope \\
2003 Jan 31 & 52670.6 & \nodata       & 17.762(0.014) & 17.397(0.015) & 17.633(0.015) & 18.069(0.027) &  CTIO 1.5-m \\

\enddata
\tablecomments{ Errors given in parenthesis, are 1$\sigma$ statistical uncertainties.}
\end{deluxetable}

\clearpage

\begin{deluxetable}{ccccccc}
\tablecolumns{7}
\tablenum{6}
\tablewidth{0pc}
\tablecaption{$JHK_s$ Photometric Sequence Around SN~2002fk.\label{IRSequence_tab}}
\tablehead{
\colhead{Star} &
\colhead{$J$}  &
\colhead{$J_{2MASS}$} &
\colhead{$H$}  &
\colhead{$H_{2MASS}$} &
\colhead{$K_s$} &
\colhead{$K_{2MASS}$}
}
\startdata

c5 & 14.599(0.014) & 14.582(0.028) & 14.259(0.008) & 14.272(0.055) & 14.148(0.009)& 14.173(0.067)\\
c8 & 15.764(0.020) & 15.712(0.062) & 15.439(0.009) & 15.564(0.101) & 15.299(0.024)& 15.408(0.179)\\
c12& 16.093(0.011) & 16.083(0.080) & 15.439(0.009) & 15.312(0.094) & 15.190(0.019)& 15.100(0.137)\\
c13& 18.224(0.064) & \nodata       & 17.628(0.077) & \nodata       & \nodata   & \nodata\\

\enddata
\tablecomments{Errors given in parenthesis, are 1$\sigma$ statistical uncertainties.}
\end{deluxetable}

\clearpage

\begin{deluxetable} {cccccc}
\tablecolumns{6}
\tablenum{7}
\tablewidth{0pc}
\tablecaption{$JHK_s$ Photometry of SN~2002fk.\label{IrPhotometry_tab}}
\tablehead{
\colhead{Date(UT)} &
\colhead{JD-2,400,000} &
\colhead{$J$} &
\colhead{$H$} &
\colhead{$K_s$} &
\colhead{Telescope} }
\startdata

2002 Sep 18 &  52535.7 &  15.162(0.011) &  15.200(0.011) &  15.107(0.011) & du Pont \\
2002 Sep 19 &  52536.7 &  14.798(0.011) &  14.859(0.011) &  14.797(0.011) & du Pont \\
2002 Sep 20 &  52537.8 &  14.540(0.011) &  14.605(0.011) &  14.532(0.015) & du Pont \\
2002 Sep 21 &  52538.7 &  14.324(0.011) &  14.405(0.011) &  14.323(0.011) & du Pont \\
2002 Sep 22 &  52539.7 &  14.132(0.011) &  14.234(0.015) &  14.103(0.015) & du Pont \\
2002 Sep 23 &  52540.7 &  14.000(0.011) &  14.133(0.011) &  13.997(0.011) & du Pont \\
2002 Sep 24 &  52541.7 &  13.860(0.016) &  14.044(0.016) &  13.863(0.017) & Baade \\
2002 Sep 24 &  52541.7 &  13.879(0.011) &  14.036(0.011) &  13.926(0.011) & du Pont \\
2002 Sep 25 &  52542.7 &  13.801(0.015) &  13.975(0.011) &  13.766(0.019) & Baade \\
2002 Sep 27 &  52544.7 &  13.764(0.015) &  13.981(0.011) &  13.741(0.017) & Baade \\
2002 Sep 29 &  52546.6 &  13.847(0.011) &  14.113(0.011) &  13.859(0.011) & du Pont \\
2002 Oct 14 &  52560.7 &  15.623(0.011) &  14.590(0.011) &  14.264(0.011) & du Pont \\
2002 Oct 15 &  52562.7 &  15.647(0.011) &  14.549(0.011) &  14.262(0.011) & du Pont \\
2002 Oct 20 &  52567.8 &  15.549(0.011) &  14.332(0.011) &  14.152(0.011) & du Pont \\
2002 Oct 23 &  52570.6 &  15.497(0.018) &  \nodata	 &  \nodata       & du Pont \\
2002 Oct 24 &  52571.8 &  15.377(0.011) &  14.175(0.011) &  14.083(0.011) & du Pont \\
2002 Nov 12 &  52590.8 &  15.854(0.011) &  14.905(0.011) &  14.922(0.011) & du Pont \\
2002 Nov 14 &  52593.8 &  16.073(0.011) &  15.030(0.011) &  15.063(0.011) & du Pont \\
2002 Nov 19 &  52597.5 &  16.367(0.011) &  15.248(0.011) &  15.249(0.011) & du Pont \\
2002 Nov 22 &  52600.6 &  16.549(0.013) &  15.367(0.014) &  15.350(0.030) & du Pont \\
2002 Dec 09 &  52617.5 &  17.713(0.039) &  16.112(0.015) &  16.010(0.027) & du Pont \\
2002 Dec 11 &  52619.7 &  17.798(0.030) &  16.185(0.011) &  16.070(0.027) & du Pont \\
2002 Dec 23 &  52631.6 &  18.534(0.042) &  16.797(0.017) &  16.549(0.033) & du Pont \\
2003 Jan 11 &  52650.6 &  19.434(0.103) &  17.597(0.027) &  17.039(0.056) & du Pont \\

\enddata
\tablecomments{Errors given in parenthesis, are 1$\sigma$ statistical uncertainties.}
\end{deluxetable}

\clearpage

\begin{deluxetable} {crrrrr}
\rotate
\tablecolumns{6}
\tablenum{8}
\tablewidth{0pc}
\tablecaption{Mean Differences among different photometric systems and our Average\tablenotemark{a} values \label{SysDiff_tab}}
\tablehead{
\colhead{} &
\colhead{$\Delta U$} &
\colhead{$\Delta B$} &
\colhead{$\Delta V$} &
\colhead{$\Delta R$} &
\colhead{$\Delta I$}
}

\startdata





$LCO/Swope - Average\tablenotemark{a}$   & $ 0.025 \pm 0.008$ & $ 0.009 \pm 0.004$ & $ 0.006 \pm 0.006$ & $ 0.006 \pm 0.004$ & $ 0.006 \pm 0.002$ \\
$CTIO/0.9m - Average\tablenotemark{a}$   & $-0.038 \pm 0.012$ & $-0.006 \pm 0.008$ & $-0.009 \pm 0.006$ & $-0.004 \pm 0.006$ & $-0.007 \pm 0.008$ \\
$CTIO/PROMPT - Average\tablenotemark{a}$ & \nodata            & $-0.006 \pm 0.028$ & $-0.003 \pm 0.010$ & $-0.008 \pm 0.012$ & $-0.008 \pm 0.008$ \\

$LOSS\tablenotemark{b} - Average\tablenotemark{a}$        & \nodata            & $-0.020 \pm 0.016$ & $-0.001 \pm 0.024$ & $ 0.004 \pm 0.026$ & $-0.070 \pm 0.016$ \\
$CfA3\tablenotemark{c} - Average\tablenotemark{a}$        & $-0.025 \pm 0.026$ & $-0.038 \pm 0.022$ & $-0.031 \pm 0.004$ & $-0.033 \pm 0.004$ & $-0.044 \pm 0.008$ \\

$CfA3\tablenotemark{c} - LOSS\tablenotemark{b}   $        & \nodata            & $-0.017 \pm 0.016$ & $-0.026 \pm 0.020$ & $-0.037 \pm 0.022$ & $ 0.022 \pm 0.022$ \\

\enddata
\tablecomments{Quoted uncertanties are weighted standard deviations}
\tablenotetext{a} {~CATS calibration (average of LCO/Swope, CTIO/0.9m, and CTIO/PROMPT measurements)}
\tablenotetext{b} {~LOSS calibration from \citet{riess09b, ganesha10}}
\tablenotetext{c} {~CfA3 calibration from \citet{hicken09}}
\end{deluxetable}






\clearpage

\begin{deluxetable} {ccc}
\tablecolumns{3}
\tablenum{9}
\tablewidth{0pc}
\tablecaption{SN~2002fk total (Galactic + host galaxy) reddening from different methods.\label{Reddening_tab}}

\tablehead{
\colhead{Method} &
\colhead{$E(B-V)$} &
\colhead{Reference} }
\startdata
$(B - V)_{Tail}$ & $0.066 \pm 0.019$ & \citet{phillips99} \\
$B_{max} - V_{max}$ & $0.009 \pm 0.044$ & \citet{phillips99} \\
$V_{max} - I_{max}$ & $0.082 \pm 0.038$ & \citet{phillips99} \\
$(V - H)$ & $0.014 \pm 0.037$ & \citet{kris04b} \\
$(V - K)$ & $0.075 \pm 0.036$ & \citet{kris04b} \\
\hline
Mean & $0.057 \pm 0.036$ &   \\
\enddata
\end{deluxetable}

\clearpage

\begin{deluxetable} {ccccccccc}
\rotate
\tablecolumns{9}
\tablenum{10}
\tablewidth{0pc}
\tablecaption{Hubble constant using different filter calibrations.\label{H0_tab}}

\tablehead{
\colhead{Filter} &
\colhead{$m_{max}$\tablenotemark{a}} &
\colhead{$M_{max}$\tablenotemark{b}} &
\colhead{$M_{max}^{1.1}$\tablenotemark{c}} &
\colhead{$\alpha$\tablenotemark{d}} &
\colhead{$\sigma_{SN}$ \tablenotemark{e}} &
\colhead{$H_0$ \tablenotemark{f}} &
\colhead{Systematic\tablenotemark{g}} &
\colhead{Reference\tablenotemark{h}} \\

\colhead{} &
\colhead{} &
\colhead{} &
\colhead{} &
\colhead{} &
\colhead{} &
\colhead{(\kmsMpc)} &
\colhead{Error} &
\colhead{}
}
\startdata
$B$ & 13.07(0.15) & -19.49(0.17) & -19.43(0.17) & -3.671(0.043) & 0.11 & $70.5^{+5.4}_{-5.0}$ & 4.5 & \citet{phillips99} \\
$V$ & 13.19(0.11) & -19.37(0.14) & -19.32(0.15) & -3.615(0.037) & 0.09 & $72.4^{+4.5}_{-4.1}$ & 4.1 & \citet{phillips99} \\
$I$ & 13.47(0.07) & -19.09(0.11) & -19.06(0.08) & -3.236(0.035) & 0.13 & $68.3^{+2.8}_{-2.7}$ & 4.9 & \citet{phillips99} \\
\hline
$J$ & 13.71(0.04) & -18.85(0.09) & -18.83(0.09) & -2.893(0.003) & 0.125 & $65.1^{+1.4}_{-1.3}$ & 4.5 & \citet{kattner12} \\
$H$ & 13.95(0.03) & -18.61(0.08) & -18.57(0.09) & -2.733(0.009) & 0.053 & $68.0^{+1.4}_{-1.2}$ & 3.2 & \citet{kattner12} \\
\hline
$J$ & 13.71(0.04) & -18.85(0.09) & -18.80(0.09) & -2.727(0.01) & 0.12 & $60.8^{+1.3}_{-1.3}$ & 4.1 & \citet{fola10} \\
$H$ & 13.95(0.03) & -18.61(0.08) & -18.59(0.09) & -2.667(0.02) & 0.16 & $65.4^{+1.2}_{-1.2}$ & 5.5 & \citet{fola10} \\
$K$ & 13.74(0.03) & -18.82(0.08) & -18.76(0.09) & -2.717(0.03) & 0.10 & $61.9^{+1.6}_{-1.5}$ & 5.4 & \citet{fola10} \\
\enddata

\tablenotetext{a}{\footnotesize Magnitude at maximum of SN~2002fk corrected for reddening (Galactic + host).}
\tablenotetext{b}{\footnotesize Peak absolute magnitude of SN~2002fk corrected for reddening (Galactic + host).}
\tablenotetext{c}{\footnotesize Peak absolute magnitude of SN~2002fk corrected for reddening (Galactic + host) and $\Delta m15$.}
\tablenotetext{d}{\footnotesize Zero point of the Hubble diagram.}
\tablenotetext{e}{\footnotesize Intrinsic dispersion in the decline rate versus luminosity relationship.}
\tablenotetext{f}{\footnotesize Statistical error includes uncertainties in apparent peak magnitude, reddening, correction for $\Delta m15$, 
and zero point of the Hubble diagram.}
\tablenotetext{g}{\footnotesize Systematic error in $H_0$ in units of \kmsMpc includes 0.08 mag in distance modulus to NGC 1309, 0.03 mag
in the photometric zero point, and the intrinsic dispersion in the decline rate versus luminosity relationship.}
\tablenotetext{h}{\footnotesize Reference for the $\Delta m 15$/$M_{max}$ relation, intrinsic dispersion in $M_{max}^{1.1}$,
and the zero point of the Hubble diagram.}
\end{deluxetable}

\clearpage

\clearpage

\clearpage







\clearpage

\clearpage

\clearpage

\clearpage

\clearpage

\clearpage

\clearpage

\clearpage

\clearpage

\clearpage

\clearpage

\clearpage

\begin{thebibliography}{}

\bibitem[Ade et al.(2014)]{ade14} Ade, P.~A.~R. et al. (Planck collaboration) 2014, \aa, in press

\bibitem[Allington-Smith et al.(1994)]{allington-smith94} Allington-Smith et~al. 1994, \pasp, 703, 983

\bibitem[Amanullah \& Goobar(2011)]{2011ApJ...735...20A} Amanullah, R., \& Goobar, A.\ 2011, \apj, 735, 20

\bibitem[Ayani \& Yamaoka(2002)]{ayani02} Ayani, K., \& Yamaoka, Y., 2002, IAUC 7976


\bibitem[Benetti et~al.(2005)]{benetti05} Benetti, S. et~al. 2005, \apj, 623, 1011

\bibitem[Benetti et~al.(2004)]{benetti04} Benetti, S. et~al. 2004, \mnras, 348, 261

\bibitem[Bessell(1990)]{bessell90} Bessell, M.~S. 1990, \pasp, 102, 1181


\bibitem[Blondin et~al.(2012)]{blondin12} Blondin, S., et~al. 2012, \apj, 143, 126

\bibitem[Branch et~al.(2003)]{branch03} Branch, D., et~al. 2003, \apj, 126, 1489

\bibitem[Bufano et~al.(2009)]{bufano09} Bufano, F., et~al. 2009, \apj, 700, 1456

\bibitem[Burns et~al.(2011)]{burns11} Burns, C.~R., et~al. 2011, \aj, 141, 19

\bibitem[Cardelli et~al.(1989)]{cardelli89} Cardelli, J.~A., Clayton, G.~C., \& Mathis, J.~S. 1989, \apj, 345, 245

\bibitem[Cartier et~al.(2011)]{cartier11} Cartier, R., et~al. 2011, \aap, 534, L15

\bibitem[Childress et~al.(2014)]{childress13} Childress, M.~J., Filippenko, A.~V., Ganeshalingam, M. \& Schmidt, B.~P. 2014, \mnras, 437, 338

\bibitem[Conley et~al.(2011)]{conley11} Conley, A., et~al. 2011, \apjs, 192, 1

\bibitem[Contreras et~al.(2010)]{contreras10} Contreras, C. et~al. 2010, \aj, 139, 519


\bibitem[Filippenko(1982)]{filippenko82} Filippenko, A.~V. 1982, PASP, 94, 715

\bibitem[Fisher et~al.(1997)]{fisher97} Fisher, A., Branch, D., Nugent, P. \& Baron, E. 1997, \apj, 481, L89

\bibitem[Folatelli et~al.(2010)]{fola10} Folatelli, G. et~al. 2010, \aj, 139, 120

\bibitem[Folatelli et~al.(2012)]{fola12} Folatelli, G. et~al. 2012, \apj, 745, 74

\bibitem[Foley(2013)]{foley13} Foley, R.~J. 2013, \mnras, 435, 273

\bibitem[Foley et~al.(2012d)]{2012ApJ...752..101F} Foley, R.~J., Simon, J.~D., Burns, C.~R., et al.\ 2012d, \apj, 752, 101

\bibitem[Foley et~al.(2012a)]{foley12a} Foley, R.~J. et~al. 2012a, \apj, 744, 38



\bibitem[F{\"o}rster et al.(2012)]{2012ApJ...754L..21F} F{\"o}rster, F., Gonz{\'a}lez-Gait{\'a}n, S., Anderson, J., et al.\ 2012, \apjl, 754, L21


\bibitem[Ganeshalingam et al.(2010)]{ganesha10} Ganeshalingam, M. et~al. 2010, \apjs, 190, 418

\bibitem[Ganeshalingam et al.(2011)]{ganesha11} Ganeshalingam, M., Li, W., Filippenko, A.~V. 2011, \mnras, 416, 2607

\bibitem[Gerardy et~al.(2004)]{gerardy04} Gerardy, C.~L. et~al. 2004, \apj, 607, 391

\bibitem[Hamuy et~al.(2009)]{hamuy09} Hamuy, M. et~al. 2009, \apj, 703, 1612

\bibitem[Hamuy et~al.(2006)]{hamuy06} Hamuy, M. et~al. 2006, \pasp, 118, 2



\bibitem[Hamuy et~al.(1991)]{hamuy91} Hamuy, M., Phillips, M.~M., Maza, J., Wischnjewsky, M., Uomoto, A., Landolt, A.~U., \& Khatwani, R. 1991, \aj, 102, 208

\bibitem[Hicken et~al.(2009)]{hicken09} Hicken, M. et~al. 2009, \apj, 700, 331

\bibitem[Hicken et~al.(2007)]{hicken07} Hicken, M. et~al. 2007, \apj, 669, L17

\bibitem[Hillebrandt \& Niemeyer(2000)]{2000ARA&A..38..191H} Hillebrandt, W., \& Niemeyer, J.~C.\ 2000, \araa, 38, 191

\bibitem[Ho et~al.(2012)]{ho12} Ho, S. et~al. 2012, \apj, 761, 14

\bibitem[Howell et~al.(2006)]{howell06} Howell, D.~A. et~al. 2006, Nature, 433, 308

\bibitem[Humphreys et~al.(2013)]{humphreys13} Humphreys, E.~.M.~L., Reid, M.~J., Moran, J.~M., Greenhill, L.~J., \& Argon, A.~L., 
2013, \apj, 775, 13

\bibitem[Jha et~al.(2006)]{jha06} Jha, S. et~al. 2006, AJ, 131, 527

\bibitem[Kattner et~al.(2012)]{kattner12} Kattner, S. et~al. 2012, \pasp, 912, 114

\bibitem[Kasen et~al.(2003)]{kasen03} Kasen, D. et~al 2003, \apj, 593, 788

\bibitem[Kasen(2006)]{kasen06} Kasen, D. 2006, \apj, 649, 939


\bibitem[Kessler et~al.(2009)]{kessler09} Kessler, R. et~al. 2009, \apjs, 185, 32

\bibitem[Koribalski et~al.(2004)]{kori04} Koribalski, B. S. et~al. 2004, \aj, 128, 16



\bibitem[Krisciunas et~al.(2001)]{kris01} Krisciunas, K. et~al. 2001, \aj, 122, 1616

\bibitem[Krisciunas et~al.(2004b)]{kris04b} Krisciunas, K. et~al. 2004b, \aj, 127, 1664

\bibitem[Krisciunas et~al.(2004c)]{kris04c} Krisciunas, K. et~al. 2004c, \aj, 128, 3034

\bibitem[Krisciunas et~al.(2003)]{kris03} Krisciunas, K. et~al. 2003, \aj, 125, 166


\bibitem[Kushida, R.(2002)]{kushida02} Kushida, R. 2002, IAUC 7973

\bibitem[Landolt(1992)]{landolt92} Landolt, A.~U. 1992, \aj, 104, 340

\bibitem[Leibundgut et~al.(1991)]{leibundgut91} Leibundgut, B. et~al. 1991, \apj, 371, L23

\bibitem[Leonard et~al.(2005)]{leonard05} Leonard, D.~C. et~al. 2005, \apj, 632, 450

\bibitem[Lira(1996)]{lira96} Lira, P. 1996, Master's thesis, U. de Chile

\bibitem[Lira et~al.(1998)]{lira98} Lira, P. et~al. 1998, \aj, 115, 234

\bibitem[Maeda et~al.(2010a)]{maeda10a} Maeda, K., Taubenberger, S., Sollerman, J., Mazzali, P., Leloudas, G., Nomoto, K., \& Motohara, K. 2010a, \apj, 708, 1703


\bibitem[Maeda et~al.(2010c)]{maeda10c} Maeda, K. et~al. 2010c, Nature, 486, 82


\bibitem[Maoz et~al.(2014)]{maoz14} Maoz, D., Mannucci, F., \& Nelemans, G. 2014, \araa, in press

\bibitem[Marion et~al.(2003)]{marion03} Marion, G.~H., H\"oflich, P., Vacca, W.~D. \& Wheeler, J.~C. 2003, \apj, 591, 316


\bibitem[Marion et~al.(2013)]{marion13} Marion, G.~H., et~al. 2013, \apj, 777, 40

\bibitem[Mazzali et~al(2005)]{mazzali05} Mazzali, P. et~al 2005, \apj, 623, L37

\bibitem[Mehta et~al.(2012)]{mehta12} Mehta, K.~T., Cuesta, A.~J., Xu, X., Eisenstein, D.~J. \& Padmanabhan, N. 2012, \mnras, 427, 2168


\bibitem[Nugent et~al.(1995)]{nugent95} Nugent, P., Phillips, M., Baron, E., Branch, D. \& Hauschildt, P. 1995, \apj, 455, L147

\bibitem[Parrent et~al.(2011)]{parrent11} Parrent, J.~T. et~al. 2011, \apj, 732, 30

\bibitem[Patat et~al.(1996)]{patat96} Patat, F., Benetti, S., Cappellaro, E., Danziger, I.~J., della Valle, M., Mazzali, P.~A. \& Turatto, M. 1996, \mnras, 278, 111

\bibitem[Patat et al.(2007)]{2007Sci...317..924P} Patat, F., Chandra, P., Chevalier, R., et al.\ 2007, Science, 317, 924

\bibitem[Persson et~al.(1998)]{persson98} Persson, S.~E., Murphy, D.~C., Krzeminski, W., Roth, M., \& Rieke, M.~J. 1998, \aj, 116, 2475

\bibitem[Persson et~al.(2002)]{persson02} Persson, S.~E., Murphy, D.~C., Gunnels, S.~M., Birk, C., Bagish, A., \& Koch, E. 2002, \aj, 124, 619

\bibitem[Perlmutter et~al.(1999)]{perl99} Perlmutter, S., et~al. 1999, \apj, 517, 565

\bibitem[Phillips et~al.(2013)]{phillips13} Phillips, M.~M., et~al. 2013, \apj, 779, 38

\bibitem[Phillips et~al.(1999)]{phillips99} Phillips, M.~M., et~al. 1999, \apj, 118, 1766

\bibitem[Phillips(1993)]{phillips93} Phillips, M.~M. 1993, \apj, 413, L105

\bibitem[Pignata et~al.(2011)]{pignata11} Pignata, G. et~al. 2011, \apj, 728, 14

\bibitem[Pignata et~al.(2008)]{pignata08} Pignata, G. et~al .2008, \mnras, 388, 971

\bibitem[Riess et~al.(2011a)]{riess11} Riess, A.~G. et al. 2011a, \apj, 730, 119

\bibitem[Riess et~al.(2009a)]{riess09a} Riess, A. G. et~al. 2009a, \apj, 699, 539

\bibitem[Riess et~al.(2009b)]{riess09b} Riess, A. G. et~al. 2009b, \apjs, 183, 109

\bibitem[Riess et~al.(2005)]{riess05} Riess, A. G. et~al. 2005, \apj, 627, 579

\bibitem[Riess et~al.(1998)]{riess98} Riess, A. G. et~al. 1998, \aj, 116, 1009

\bibitem[R{\"o}pke et al.(2012)]{2012ApJ...750L..19R} R{\"o}pke, F.~K., Kromer, M., Seitenzahl, I.~R., et al.\ 2012, \apjl, 750, L19

\bibitem[Scalzo et~al.(2010)]{scalzo10} Scalzo, R.~A. et~al. 2010, \apj, 713, 1073

\bibitem[Schlafly \& Finkbeiner(2011)]{schlafly11} Schlafly, E.~F. \& Finkbeiner, D.~P. 2011, \apj, 737, 103

\bibitem[Schlegel et~al.(1998)]{schlegel98} Schlegel, D.~J., Finkbeiner, D.~P. \& Davis, M. 1998, \apj, 500, 525

\bibitem[Skrutskie et~al.(2006)]{skrutskie06} Skrutskie, M.~F., et~at. 2006, \aj, 131, 1163

\bibitem[Silverman et~al.(2010)]{silverman10} Silverman, J.~M., et~al. 2010, \mnras, 410, 485

\bibitem[Silverman et~al.(2012a)]{silverman12a} Silverman, J.~M., et~al. 2012, \mnras, 425, 1789

\bibitem[Silverman \& Filippenko(2012b)]{silverman12b} Silverman, J.~M. \& Filippenko, A.~V. 2012, \mnras, 425, 1917

\bibitem[Silverman et~al.(2013)]{silverman13} Silverman, J.~M., Ganeshalingam, M. \& Filippenko, A.~V. 2013, \mnras, 430, 1030

\bibitem[Simon et al.(2009)]{2009ApJ...702.1157S} Simon, J.~D., Gal-Yam, A., Gnat, O., et al.\ 2009, \apj, 702, 1157

\bibitem[Stanishev et~al.(2007)]{stanishev07} Stanishev, V., et~al. 2007, \aap, 469, 645

\bibitem[Sternberg et al.(2011)]{2011Sci...333..856S} Sternberg, A., Gal-Yam, A., Simon, J.~D., et al.\ 2011, Science, 333, 856

\bibitem[Stritzinger et~al.(2002)]{stritzinger02} Stritzinger, M. et~al. 2002, \aj, 124, 2100

\bibitem[Stritzinger et~al.(2010)]{stritzinger10} Stritzinger, M. et~al. 2010, \apj, 140, 2036


\bibitem[Sullivan et al.(2011)]{sullivan11} Sullivan, M. et~al. 2011, \apj, 737, 102

\bibitem[Tanaka et~al.(2008)]{tanaka08} Tanaka, M. et~al. 2008, \apj, 677, 448


\bibitem[Thomas et~al.(2004)]{thomas04} Thomas, R.~C. et~al. 2004, \apj, 601, 1019

\bibitem[Thomas et~al.(2007)]{thomas07} Thomas, R.~C. et~al. 2007, \apj, 654, L53

\bibitem[Thomas et~al.(2011a)]{thomas11a} Thomas, R~C. et~al. 2011a, \pasp, 123, 237

\bibitem[Thomas et~al.(2011b)]{thomas11b} Thomas, R.~C. et~al. 2011b, \apj, 743, 27

\bibitem[Timmes et~al(2003)]{timmes03} Timmes, F.~X., Brown, F.~E., \& Truran, J.~W. 2003, \apj, 590, L83

\bibitem[Wang \& Wheeler(2008)]{wangwheeler08} Wang, L. \& Wheeler, J.~C. 2008, \araa, 46, 433


\bibitem[Wang \& Qiu(2002)]{wang02} Wang, J.,\& Qiu, Y.L. 2002, IAUC 77973

\bibitem[Wang et~al.(2009)]{wang09} Wang, X., et~al. 2009, \apj, 697, 380

\bibitem[Wood-Vasey et~al.(2008)]{wood08} Wood-Vasey, W. M. et~al. 2008, \apj, 689, 377

\bibitem[Yamanaka et~al.(2009)]{yamanaka09} Yamanaka, M. et~al. 2009, \apj, 707, L118


\bibitem[Zelaya et~al.(2014)]{zelaya13} Zelaya et~al. 2014, (in preparation)

\end{thebibliography}
\end{document}